\documentclass[prb,aps,twocolumn,superscriptaddress]{revtex4-1}

\usepackage{amsmath}
\usepackage{feynmp-auto}
\usepackage{graphicx}
\usepackage{grffile}
\graphicspath{{images/}}
\graphicspath{{Figures/}}

\begin{document}
\title{Exact equilibrium results in the Interacting Resonant Level Model }

\author{Gonzalo Camacho}
\affiliation{School of Physical Sciences, University of Kent, Canterbury CT2 7NH, United Kingdom}

\author{Peter Schmitteckert}
\affiliation{Institute for Theoretical Physics and Astrophysics, Julius-Maximilian University of W\"urzburg, Am Hubland, 97074 W\"urzburg, Germany}

\author{Sam T.~Carr}
\affiliation{School of Physical Sciences, University of Kent, Canterbury CT2 7NH, United Kingdom}

\date{\today}

\begin{abstract}
We present exact results for the susceptibility of the interacting resonant level model in equilibrium. Detailed simulations using both the Numerical Renormalization Group and Density Matrix Renormalization Group were performed in order to compare with closed analytical expressions.  By first bosonizing the model and then utilizing the integrability of the resulting boundary sine-Gordon model, one finds an analytic expression for the relevant energy scale $T_K$ with excellent agreement to the numerical results.  On the other hand, direct application of the Bethe ansatz of the interacting resonant level mode does not correctly reproduce $T_K$ -- however if the bare parameters in the model are renormalised, then quantities obtained via the direct Bethe ansatz such as the occupation of the resonant level as a function of the local chemical potential do match the numerical results.  The case of one lead is studied in the most detail, with many results also extending to multiple leads, although there still remain open questions in this case.
\end{abstract}
\maketitle

\section{Introduction}
The Interacting Resonant Level Model (IRLM) is one of the most important models in the study of the effects of quantum impurities in one dimensional systems. Although most of the recent research has focused on non-equilibrium properties\cite{Andrei,*[{Erratum: }]AndreiErratum,Vinkler,Borda1,SchillerAndrei,Boulat1,Boulat2,BBSS,Sampaper,Sampaper2,Sampaper3,vonDelftIRLM2018} of the model, its equilibrium properties still prove to have interest on their own. The relation of the model with the Anisotropic Kondo Model\cite{WiegmannFinkelshtein,Schlottman} and the two Ohmic state system \cite{Ohmic} (known as the Spin Boson model), as well as different variations like the coupling of the impurity to a Luttinger liquid,\cite{Berkovits1,Berkovits2,Rylands} make the IRLM fundamental in the study of such strongly correlated systems.

 Originally, the IRLM was derived from the anisotropic Kondo model in the work by Wiegmann and Finkelshtein,\cite{WiegmannFinkelshtein} where the partition functions of both models are shown to be equivalent. Thermodynamic quantities like the specific heat and the charge susceptibility can be found there, proving Wilson's ratio relation; however, these results are obtained by perturbation theory, which accounts only for a small region in parameter space. More recent works\cite{Borda1,Borda2,Kiss1} have applied perturbative RG in the model, in the same line as the work developed by Schlottman.\cite{Schlottman} As interaction is increased however, there are marked deviations between numerical results and analytical predictions,\cite{Kiss1,Borda1} therefore demonstrating that perturbative RG is not suitable to describe physical properties of the model beyond the weak coupling limit.
 
 In the strong coupling regime, where interaction with the impurity is bigger than the bandwidth, bosonization has provided generic results in the model,\cite{Borda2} along wih the Anderson-Yuval approach,\cite{AY,Fabrizio}  commonly used to treat the Kondo model in analogy with the X-ray threshold problem. These methods allow for the exact calculation of thermodynamic exponents in the theory; a numerical confirmation of these results is one of the purposes of the work presented here.
 In addition, the above mentioned methods do not allow for the calculation of an exact expression of the relevant energy scale $T_{K}$ appearing in the model (the analog to the \emph{Kondo temperature} in the Kondo model), meaning such expressions are approximate.
It has been seen, for example, that bosonization fails to reproduce the Numerical Renormalization Group (NRG) data points for this relevant energy scale in the single channel version of the model.\cite{Borda2}  

There is a simple explanation for this disagreement: the energy scale $T_K$ may be expressed as $T_K \sim C\, (t')^\alpha$, where $t'$ is the hybridisation between the impurity and the lead(s), and $\alpha$ is the interaction dependent exponent (see expressions \eqref{energyscale1} and \eqref{TKexact} later for a more precise definition). What hasn't been fully appreciated in previous work is that the prefactor, $C$, also depends on the interaction parameter in a non trivial way, meaning that to directly compare the analytic result to numerical ones, one must either numerically extract the exponent $\alpha$, or analytically calculate the prefactor $C$.  Both of these are done in the present work; and that this prefactor can be obtained from the appropiate field theory is one of the main points we illustrate here.

 Although calculation of such exact expressions have been pioneered by the use of Bethe ansatz methods,\cite{AndreiKondo,TsvelickWiegmann} and in particular, exact expressions have been calculated recently in the IRLM when the impurity site is attached to a Luttinger liquid\cite{Rylands} as well as for the multichannel version of the model,\cite{Ponomarenko} this is not the approach we shall follow here. We claim that this prefactor can be obtained in an exact way by identifying the corresponding low-energy field theory which in this case is a boundary sine-Gordon model, and then exploiting the integrability of this low energy field theory.  This provides an expression for $T_K$ that (for the one-channel case) is in excellent agreement with numerics on a lattice, as well as agreeing with strong coupling expansions to leading order in the inverse of the interaction strength.
  
 Simulations have been performed using both the Density Matrix Renormalization Group (DMRG) and Numerical Renormalization Group (NRG) techniques, in order to obtain the exponent $\alpha$ of the theory for the cases of one, two, three and four channels. Both the analytic formula and numerical results prove to be in very  good agreement. The relevant energy scale $T_{K}$ in the single lead version of the model is also calculated numerically and compared with exact expressions from the field theory. The philosophy behind this approach is to open the door to calculate such exact expressions in the model by extrapolation of the method to the multichannel IRLM, as well as to show that both the lattice and the field theory approaches must give identical low-energy descriptions. The motivation to work on this assertion comes from previously reported results in the two channel version of the model, where it has been shown that there is apparent conflict between a lattice and a field theory description in the strong coupling limit.\cite{SchillerAndrei} A strong coupling expansion in the lattice breaks the duality $U\leftrightarrow 1/U$ in the model,\cite{SchillerAndrei} whereas in a field theoretic description it is claimed that such disagreements can be fixed if one chooses the appropiate regularization for the fields in the continuum. The results we are about to show here do not support the idea that regularization schemes in the continuum are relevant for describing low energy processes of the model; instead, both lattice and field descriptions of the model must provide the same answers when it comes to describe low-energy properties.
 
 The model presents quantum critical points at finite values of interaction. To discuss the nature of quantum phase transitions, a good understanding of the strong coupling limit is needed, in order to build up an effective low energy theory. We show this low-energy description of the strong coupling limit to be essential to interpret the change in the ground-state of the system when the quantum phase transition occurs.

The paper is structured as follows. In section II, we introduce the model in its multichannel version (MIRLM). In this section we also review some thermodynamic concepts and give a precise definition of the thermodynamic energy scale $T_K$ that will be used in subsequent calculations. In section III, we study the single channel version of the IRLM. We bosonize the model and map it to a generalised Boundary sine Gordon model (BsG), in order to calculate in an exact way the scaling exponent of the coupling parameter. Further correspondence of the IRLM and the BsG is explored by considering exact expressions of the relevant energy scale $T_{K}$ of the BsG from field theory. Analytic formulas are then compared with numerical results given by Density Matrix Renormalization Group (DMRG) and Numerical Renormalization Group (NRG) in the IRLM. The strong and weak coupling limits of the theory are confirmed both analytically and numerically. In section IV, numerical simulations are performed to compare with closed analytical results of the dot occupation\cite{Rylands} obtained by the Bethe-ansatz; we notice this agreement between numerics and analytic formulas holds under a proper fixing of the exponent. We emphasize here the importance of appropiate scaling and the role played by the relevant energy scale $T_{K}$. In section V, we demonstrate an important consequence of our results: a recalculation of the Toulouse point where the Anisotropic Kondo Model maps to the Resonant Level Model ($U=0$). We show this value to differ from the one usually given in the literature.\cite{Gogolin} Section VI is then dedicated to the extension of field theoretic techniques for the multichannel case, giving the exact exponent for the thermodynamic scale, as well as a discussion of the quantum phase transitions of the model by low-energy effective theory arguments, thus studying the strong coupling limit in the multichannel case. In addition, the integrability of the model and its exact solution reported in previous works\cite{Rylands,FilyovW,Ponomarenko,TsvelickWiegmann} is discussed along with the results obtained.

\section{The model}
 
 \begin{figure}
\begin{center}
\includegraphics[width=3in,clip=true]{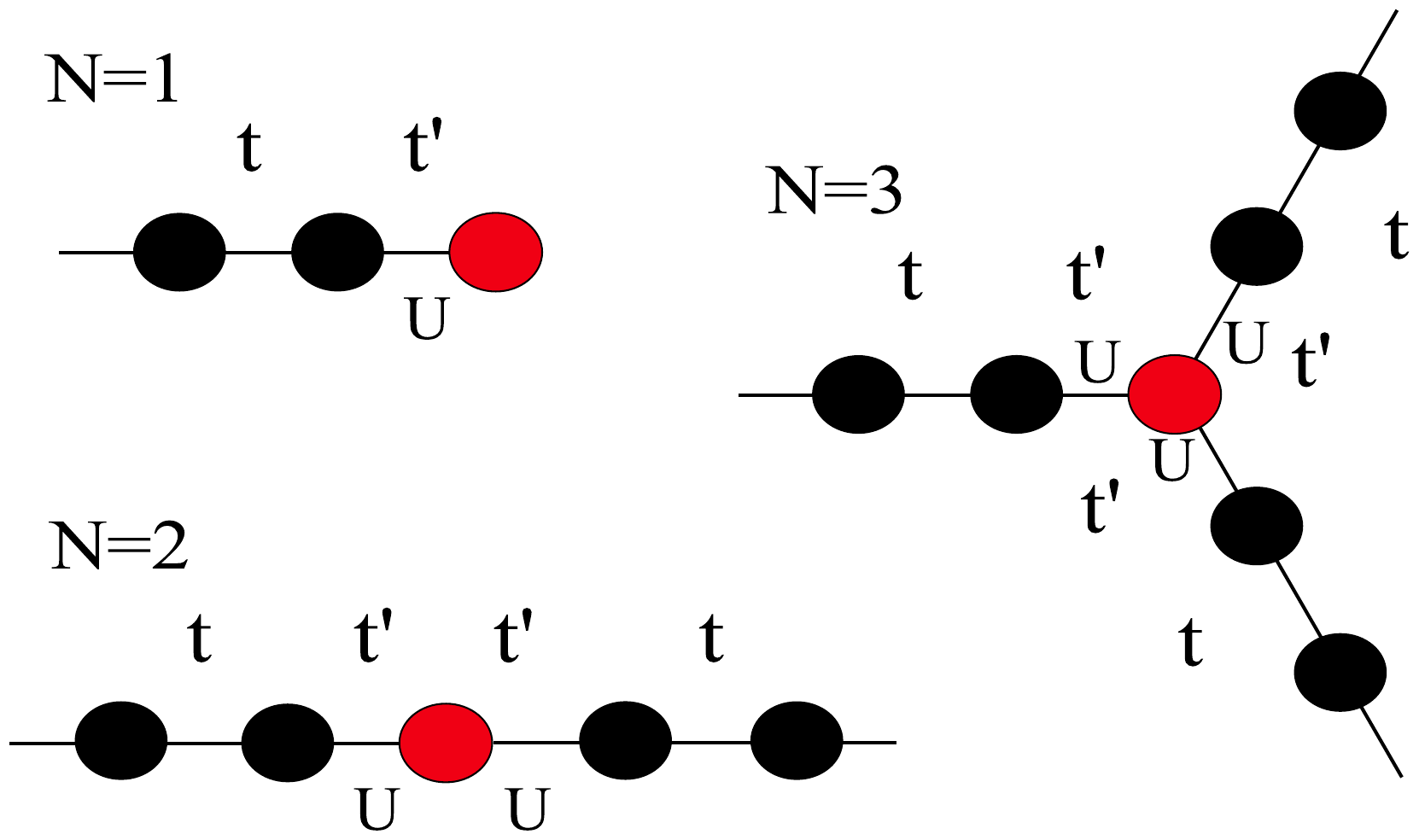}
\end{center}
\caption{Interacting Resonant Level Model (IRLM) for single, two and three channel cases. The red circle represents the impurity site, whereas black dots represent sites on the leads. The hopping parameter $t$ is related with the lead density of states, whereas $t'$ represents a hybridization between the dot and the leads. An interaction $U$ betweeen the last site of the lead and the dot is present for each channel. The leads extend to infinity.}
\label{fig:model}
\end{figure}
 
The $N$-channel IRLM comprises of a single quantum impurity site hybridised with $N$ semi-infinite non-interacting leads which are conveniently modelled as tight-binding chains.  The fermions are spinless so each site has a maximum occupation of one, and the only interaction is between the edge of each wire and the impurity site.  The model is illustrated schematically in Fig.~\ref{fig:model} for the cases $N=1,2,3$.  In principle, the hybridization strength $t'_\gamma$ and interaction $U_\gamma$ could depend on the lead index $\gamma$, but for simplicity we limit ourselves to the case where all leads are identical.
 
 In second quantized notation, the lattice version of the model is described by the following Hamiltonian:
 \begin{eqnarray}\label{Hlattice}
 H=H_{0}+\varepsilon_{0}d^{\dagger}d + t'\sum_{\gamma=1}^{N}\bigg( d^{\dagger}c_{0,\gamma} +{\text{h.c}}\bigg)\nonumber\\ 
 + U\sum_{\gamma=1}^{N}\bigg(d^{\dagger}d -\frac{1}{2}\bigg)\bigg(c_{0,\gamma}^{\dagger}c_{0,\gamma}-\frac{1}{2}\bigg) 
 \end{eqnarray}
 where the $d$ and $d^\dagger$ operators represent fermionic annihilation and creation operators on the impurity site, while the $c_{0,\gamma}$ operators refer to fermions at the end of the wire for the channel $\gamma$. The parameter $\varepsilon_{0}$ represents a local chemical potential on the impurity site, $t'$ the hybridization parameter between the channels and the impurity, and $U$ is the interaction parameter between the impurity and the leads.  The minus one half terms in the interaction are the ground state expectation values $\langle d^\dagger d \rangle$ and $\langle c^\dagger_0 c_0\rangle $ respectively at $\varepsilon_0=0$.  This ensures that $\varepsilon_0=0$ is the resonance point, and we have implicitly made the assumption that the tight-binding leads are half-filled.
  
 The non-interacting part of the Hamiltonian has the usual tight-binding form for non-interacting fermions:
 \begin{eqnarray}
 H_{0}=  -t   \sum_{\gamma=1}^{N}  \sum_{i=0}^{+\infty}  c_{i+1}^{\dagger}c_{i}+\text{h.c}
 \label{eq:H0lattice}
 \end{eqnarray}
The tight-binding form of the leads is particularly convenient for both numerical work and strong coupling analysis, however we will also be using field-theory and Bethe ansatz results which consider continuum leads with a constant density of states $\nu$.  As is usually the case, the low-energy properties of the different models of the leads coincide so long as the hybridization is much less than the band-width $t'/t \ll 1$, although importantly we will put no such restriction on $U$.  We will come back to this point repeatedly as it is one of the key results of this work that not only do the universal properties match in the two cases, but also \textit{non-universal properties} such as the exact resonance width and line shape; and we will derive exact relationships between parameters for the cases of tight-binding and continuum leads.  This is very important as the model is only integrable (exactly-solvable) for continuum leads, while numerical work is much more convenient on tight-binding chains.

Let us now look at the basic properties of the model, where for clarity we will restrict ourselves to the one-lead case, $N=1$.  We will return to the multichannel case in Sec.~\ref{sec:Nbiggerthan1}.  The model is characterised by four parameters: the hybridization $t'$, the interaction $U$, the local chemical potential at the impurity site, $\varepsilon_{0}$, and the density of states in the leads $\nu$.  For the case of semi-infinite, half-filled, tight-binding leads, $\nu=1/\pi t$, however by writing everything in terms of $\nu$, the results are universal for any regularisation of the (non-interacting) leads, so long as the density of states $\nu$ can be approximated to be energy independent at low energies.

There are two different limits to the model which are trivially solvable -- the non-interacting limit $U=0$ and the decoupled limit $t'=0$.  We now look at each of these in turn.

\subsection{The non-interacting case $U=0$}

When there is no-interaction, the impurity level gains a width $T_0$ given by
\begin{eqnarray}\label{non_interacting_T}
 T_{0}=\pi\nu (t')^{2}
 \end{eqnarray}
due to the hybridization with the lead.  This means that the occupation of the level $n_d= \langle d^\dagger d \rangle$ as a function of its energy $\varepsilon_0$ is given by the well known textbook form,
\begin{eqnarray}\label{nd-U0}
n_{d}=\frac{1}{2}-\frac{1}{\pi}\arctan\bigg(\frac{\varepsilon_{0}}{T_{0}}\bigg).
\end{eqnarray}
Upon differentiation, this gives the standard Lorenzian shape for the local charge susceptibility, or equivalently in the non-interacting case, the impurity density of states.  The single energy scale of the resonance width $T_0$ governs all of the properties of the system.

When the interaction is switched on $U\ne 0$, two things happen: firstly, the occupation no longer has the exact form Eq.~(\ref{nd-U0}), and secondly the resonance width $T_0$ is renormalised.  In analogy with the Kondo model, we will call this renormalised width $T_K$, where we will often compare to the non-interacting case $T_0=T_K(U=0)$.  Furthermore however, if we want to make precise comparisons between theory and numerics, we need to define the width without regard to a particular line-shape.  We therefore define this energy scale in terms of the local charge susceptibility at resonance:
 \begin{eqnarray}\label{Gammades}
 T_{K}^{-1} =-\pi\bigg(\frac{\partial n_{d}}{\partial \varepsilon_{0}}\bigg)\bigg|_{\varepsilon_{0}=0}.
 \end{eqnarray}

\subsection{The decoupled case $t'=0$}

There is another trivially exactly solvable limit of the IRLM, where the interaction $U$ is non-zero, however the impurity site is decoupled from the lead $t'=0$.
In this case, because there is no hybridization between particles from the leads and the dot, the occupation number on the dot cannot change and is a good quantum number, and hence this problem is effectively non-interacting as well. Fermions coming from the lead to the edge of the wire will encounter a localised potential that can be represented as a delta barrier of amplitude:
\begin{eqnarray}
V(x)=\pm \frac{U}{2} \delta(x)
\label{eq:potscat}
\end{eqnarray}
therefore reducing the N-body problem to a single-particle scattering problem. Particles scattered by this potential will experience a phase shift $\pm \delta$ on the wave-function. This phase-shift is given by\cite{SchillerAndrei,Borda2,Kiss1,delta_potential}
\begin{eqnarray}\label{phase}
\delta = \arctan\bigg(\frac{U\pi\nu}{2}\bigg)
\end{eqnarray}
where $\nu$ represents the density of states on the leads as usual.  It is worth emphasising that as this is a problem of single-particle quantum mechanics, this scattering phase shift may be calculated for any particular model of the leads -- if one considers a field theoretic (continuous) model of the leads, ones finds $\delta=\pi U\nu /2$, while the result above corresponds to tight binding leads.  As an aside, it is curious to note that even for continuous leads, there are two ways of regularising the delta function barrier, one of which gives the usual result $\delta \propto U\nu$ and the other of which gives the same as the tight-binding model. Eq.~\eqref{phase} above.\cite{delta_potential}  However the main message is that if one takes the scattering phase shift $\delta$ (more precisely, the scattering phase shift at the Fermi-energy when $t'=0$) as the model parameter rather than the (bare) interaction strength $U$, then the results are independent of details of the lead itself.  This is in spirit the same as using the density of states on the leads $\nu$ (which is independent of details of the lead) rather than the hopping parameter $t$ which implies a specific model of the lead.

Let us now define the following dimensionless interaction parameter:
\begin{eqnarray}\label{defg}
g=\frac{2\delta}{\pi}
\end{eqnarray}
associated with the scattering phase shift. This parameter will prove to be convenient for later purposes, and it is restricted to lie between values $[-1,+1]$ for all possible values of interaction $U$.  For weak interactions much less than the bandwidth $U \ll t$, we have $g \approx U\nu$, and many authors have concentrated on this perturbative regime.\cite{Borda1,Schlottman,Kiss1}  However, we will show that leaving $g$ as the input to more advanced field-theoretic techniques rather than the bare $U$ removes this restriction that $U$ is small, and allows numerics to be compared to analytic results over the whole range of interactions.  It is also worth stressing that solving for $g$ in terms of $U$ is a one-dimensional quantum mechanics scattering problem, so $g$ is not a phenomenological parameter, but something that can be calculated exactly in terms of microscopic parameters of the theory, even on a lattice.  It is this feature of the model that allows us to obtain exact expressions from field theory that precisely match numerical simulations on a lattice.


\section{Exact results for the resonance width for $N=1$ }
\label{sec:N1}


To summarise the previous section which sets the mood for the following: there are still four parameters that govern the model: (i) the hybridisation $t'$, which can also be expressed in terms of the non-interacting resonance width $T_0$; (ii) the interaction $U$, which is expressed through the scattering phase shift it gives $g$ defined in Eq.~\eqref{defg} above; (iii) the density of states on the lead $\nu$; and (iv) the impurity energy $\varepsilon_0$.    
We now proceed to discuss what happens when both the hybridisation $T_0$ and the interaction $g$ are non-zero, where we will discuss first the renormalisation of the level width, and then go on to look at the exact resonance line-shape.  In all cases, we calculate the property analytically and then compare with numerical results.

\subsection{Bosonization}
\label{sec:bosonization}

We begin the analysis of the thermodynamic properties of the IRLM by using the technique of Bosonization.\cite{Gogolin,GiamarchiBook}  This will serve two purposes: firstly, it will allow us to calculate the scaling exponent $\alpha$ relating the resonance width to the bare hybridization $T_K \sim (t')^\alpha$, where we will demonstrate numerically that the result is exact; and secondly it will demonstrate the mapping from the IRLM on to the boundary sine-Gordon model, which will allow us to use the integrability of the latter model in the next session in order to get an exact expression for the resonance width.  While this bosonization calculation has been done many times before,\cite{Gogolin,GiamarchiBook,Borda2} we find it useful to repeat it here.

To apply bosonization, we first need to rewrite the lattice Hamiltonian (\ref{Hlattice}) in the continuum.   The field theoretic approach replaces lattice operators $c_{i}$ by fermionic fields $\psi(x)$ which results (for $N=1$) in\cite{Gogolin,Andrei,SchillerAndrei}
\begin{eqnarray}\label{H_single_channel}
H=-i\int_{-\infty}^{+\infty}dx\big(\psi^{\dagger}(x)\partial_{x}\psi(x)\big)+\varepsilon_{0}d^{\dagger}d  \nonumber\\
+t'(\psi^{\dagger}(0)d + \text{h.c})
+U:\psi^{\dagger}(0)\psi(0):\bigg(d^{\dagger}d-\frac{1}{2}\bigg),
\end{eqnarray}
where $:\psi^{\dagger}(0)\psi(0):=\psi^{\dagger}(0)\psi(0)-\langle \psi^{\dagger}(0)\psi(0) \rangle$ is the normal ordering of the fermionic field that ensures that the resonance is at $\epsilon_0=0$ (c.f. the discussion under Eq.~\eqref{Hlattice}). Two things have been combined in this mapping -- firstly, the spectrum in the leads has been linearized in the vicinity of the Fermi points, with the operators normalised in a way to set the Fermi velocity $v_{F}=1$.  Secondly, the lead has been unfolded\cite{Gogolin}.  This basically means that the original model with operators near two Fermi points -- which for the sake of argument we will call left (L) and right (R) -- on the semi infinite lead have been replaced for a model with only right movers on the whole real line: the left movers for $x<0$ are now written as right movers for $x>0$, which matches the boundary condition of the semi-infinite lead at $x=0$.\cite{Gogolin}  It is also worth pointing out here that in making the continuum limit, we have implicitly supposed that there is a high-energy cutoff $D$, which may qualitatively be thought of as the band-width, and plays the role of the (inverse of the) density of states $\nu$ in the lattice model.  This means that even though we have set $v_F=1$, we still have the same number of degrees of freedom in parameter space as in the lattice model.

We now apply the bosonization procedure\cite{Gogolin,GiamarchiBook} to the continuous Hamiltonian above.
The (right-moving) fermionic field operators are expressed in terms of a chiral bosonic field $\phi(x)$ by the relation:
\begin{eqnarray}\label{bosonization_relation}
\psi(x)=\frac{\eta}{\sqrt{2\pi}}e^{i\sqrt{4\pi}\phi(x)},
\end{eqnarray}
where $\eta$ is the Klein factor that ensures anticommutation rules between different species of fermions to be satisfied. 
It is convenient also to rewrite the $d$ operators on the impurity site in terms of a spin-$1/2$ operator, $\vec{S}$:
\begin{eqnarray}
S^{z}&=&d^{\dagger}d-\frac{1}{2}\nonumber\\
\eta_0 S^{+}&=&d^{\dagger}
\end{eqnarray}
where $\eta_0$ is again a Klein factor.
Applying this transformation, the bosonized version of the one channel IRLM then reads:
\begin{eqnarray}
H=H_{0}+\frac{t'}{\sqrt{2\pi}}\big(\eta_0 \eta e^{-i\sqrt{4\pi}\phi(0)}S^{-}+\text{h.c}\big)+\varepsilon_0 S^z
\label{hambos}
\end{eqnarray}
where the non-interacting (quadratic) part of the hamiltonian is given by:
\begin{eqnarray}
H_{0}=\frac{1}{2}\int_{-\infty}^{+\infty} dx(\partial_{x}\phi(x))^{2}+\frac{U}{\sqrt{\pi}}\partial_{x}\phi(0)S^{z}
\label{eq:H0bos1}
\end{eqnarray}
The combination of Klein factors $\eta_0 \eta$ is a constant of motion, which we will choose to be $+1$ and so will not consider these further.\cite{Gogolin}  It is worth explicitly pointing out that bosonization has made the interaction term $U$ quadratic so it can be treated exactly at the expense of a more complicated form for the hybridization term $t'$.  For later convenience, the local chemical potential term, $\varepsilon_0$ is not included in $H_0$, and in future will be excluded from the Hamiltonian when we are focussed only on resonance $\varepsilon_0=0$.

Let us now consider the Hamiltonian \eqref{hambos} in the limit $t'=\varepsilon_{0}=0$.  As discussed previously, this limit corresponds to a scattering problem, which on solving the model \eqref{eq:H0bos1} corresponds to scattering at $x=0$ with a phase shift of $\delta = \pm U S^z$.  This agrees with the lattice case \eqref{phase} only in the limit $U\nu \ll 1$ -- however one can make these two models agree for all $U$ by rewriting the bosonized continuous Hamiltonian as
\begin{eqnarray}
H_{0}=\frac{1}{2}\int_{-\infty}^{+\infty} dx(\partial_{x}\phi(x))^{2}+\sqrt{\pi} g \; \partial_{x}\phi(0)S^{z}
\label{eq:H0bos}
\end{eqnarray}
where $g=2\delta/\pi$ as defined in \eqref{defg}.  This is often known as the \textit{phase shift substitution}, but ultimately its goal is to match the low-energy physical properties of the bosonized model (which depend on its high-energy regularisation) with the original microscopic model (\ref{eq:potscat}) which in general may have a different regularisation (e.g. be on a lattice) from the standard one used in bosonization.\cite{Kiss1,Borda2,ZCJA,SchillerAndrei,Fabrizio}  A similar procedure can be used to compute the ultra-violet cutoff $D$ that should be used in the bosonized theory to match the low-energy physics of the original model; this calculation is however not required here and will be deferred to section \ref{sec:prefactor}.

By applying a unitary transformation to the Hamiltonian (\ref{hambos})
\begin{eqnarray}
\bar{H}=\mathcal{U}^{\dagger}H\mathcal{U}
\end{eqnarray}
with
\begin{eqnarray}\label{unitary_transf}
\mathcal{U}=e^{i\sqrt{4\pi} g S^{z}\phi(0)}
\end{eqnarray}
we find that we have eliminated the interaction term from the Hamiltonian\cite{Gogolin}
\begin{eqnarray}
\bar{H}=\bar{H}_{0}+\frac{t'}{\sqrt{2\pi}}\big(S^{-}e^{-i\sqrt{4\pi}(1-g)\phi(0)}+\text{h.c}\big)
\label{eq:vertex}
\end{eqnarray}
where now
\begin{eqnarray}
\bar{H}_{0}=\frac{1}{2}\int_{-\infty}^{+\infty} dx(\partial_{x}\phi(x))^{2}
\end{eqnarray}
is the Hamiltonian of a standard non-interacting chiral boson.

The above hamiltonian Eq.~\eqref{eq:vertex} is identical to the bosonized version of the anistropic Kondo model,\cite{WiegmannFinkelshtein,Gogolin} which as we have already noted is historically how the IRLM was first introduced.  We will return to the relationship to the Kondo model in Section \ref{sec:Kondo}.  For now, we can apply very standard analysis to the above Hamiltonian.\cite{Gogolin}  A vertex operator $e^{i \beta \phi /\sqrt{2} }$ has scaling dimension $d=\beta^2/16\pi$ (see e.g. Ref.~\onlinecite{Gogolin}), and note that the scaling dimension should not be confused with the dot annihilation operator; although both are denoted by $d$, the context should make clear which one is being referred to.  Hence the scaling dimension $d$ of the vertex operator in Eq.~\eqref{eq:vertex} is:
\begin{eqnarray}\label{scaling}
d=\frac{(1-g)^{2}}{2}
\end{eqnarray}
For $d<1$ (the space-time dimension of the boundary), this is a relevant operator, and therefore gives rise to an emergent low energy scale
\begin{eqnarray}\label{exp_asymptotic}
\frac{T_{K}}{D}\sim\bigg(\frac{t'}{D}\bigg)^{\alpha}
\end{eqnarray}
where $D$ is introduced as the high-energy cutoff of the field theory, and the exponent
\begin{eqnarray}\label{alpha}
\alpha=\frac{1}{1-d}=\frac{2}{1+2g-g^{2}}
\end{eqnarray}
This can be derived by a number of standard techniques, such as the renormalisation group or self-consistent mean field theory.\cite{GiamarchiBook,Gogolin}

This exponent has been calculated many times before,\cite{Borda1,Borda2,Kiss1,Vinkler} albeit sometimes without the $g^2$ term and sometimes directly associating $g$ with the interaction strength $U$ rather than with the scattering phase shift $g=2\delta/\pi$.  There has therefore been a lot of discussion in the literature as to whether this result for the exponent (\ref{alpha}) is exact, or is a perturbative expansion in small interaction strength.  We will show by comparing to numerics in Sec.~\ref{sec:numerics} and a strong coupling expansion in Sec.\ref{sec:strong} that the result is in fact non-perturbative in interaction and is exact for any interaction strength.  Before getting to this however, let us see if we can say anything more precise than the asymptotic relationship in Eq.~\eqref{exp_asymptotic}

\subsection{The boundary sine-Gordon model and the prefactor $f(\alpha)$}
\label{sec:prefactor}

Equation  \eqref{exp_asymptotic} only gives an asymptotic relationship between the energy scale and the hybridization parameter $t'$. In particular, it says nothing about the proportionality factor of the expression. 
Let us therefore write instead
\begin{eqnarray}\label{energyscale1}
\frac{T_{K}}{D}=\tilde{f}(\alpha)\bigg(\frac{t'}{D}\bigg)^{\alpha} + \mathrm{\ higher\ order\ terms\ in\ }t',
\end{eqnarray}
where the pre-factor $\tilde{f}(\alpha)$ is is a function of the interaction strength $g$ which is conveniently expressed through the exponent $\alpha$.  It is worth pointing out that while not perturbative in interaction strength, our asymptotic scaling relation Eq.~\eqref{exp_asymptotic} is formally the leading term in a perturbation expansion in the hybridisation strength $t'$.  This is not an important issue, as one must have $T_K \ll D$ and hence $t' \ll D$ anyway in order for the field theory to quantitatively describe a lattice model -- the bigger surprise in this model perhaps is that the interaction strength $U$ is not subject to this limit.

Nonetheless, there is no reason to suppose that $\tilde{f}(\alpha)$ is independent of interaction strength, and therefore needs to be included if one wants to directly compare emergent energy scales $T_K$ from numerics with theoretical predictions.  While one can get around this by processing the numerical results for different values of $t'$ to obtain the exponent $\alpha$, which we will demonstrate in the next section, in this section we will show that we can also obtain an exact expression for the pre-factor $\tilde{f}(\alpha)$.

The Hamiltonian \eqref{eq:vertex} may be rewritten
\begin{multline}
\bar{H}=\bar{H}_{0}+\sqrt{\frac{2}{\pi}}t' \left[ S^x \cos\left( \sqrt{4\pi} (1-g) \phi(0) \right) \right. \\
\left. - S^y \sin \left( \sqrt{4\pi} (1-g) \phi(0) \right) \right]
\end{multline}
Let us make the assumption that in the ground state at temperature $T=0$, one can rotate the spin-quantisation axis so the spin points in the $x$ direction, where either of the $S^x = \pm 1/2$ eigenvalues can be taken and the hamiltonian above can be written as a boundary sine-Gordon model, with action described by:
\begin{eqnarray}
H=H_0 + \frac{t'}{\sqrt{2\pi}} \cos\big(\frac{\beta}{\sqrt{2}}\phi(0)\big)
\label{eq:bsG}
\end{eqnarray}
where $\beta^{2}= 8\pi (1-g)^2$. The $\sqrt{2}$ factor inside the cosine is a standard convention in the literature on the boundary sine-Gordon model relating to the fact that it remains integrable when a bulk $\cos \beta \phi$ term is added, which is not present in this case.  We stress that we know of no good \textit{a-priori} justification that this assumption that decouples the impurity dynamics from the lead is a good one.  However in sections \ref{sec:numerics} and \ref{sec:strong} we will show that the results following this assumption are in very good agreement with numerics and the strong coupling limit, and we will also comment again on this assumption when we discuss the direct Bethe-Ansatz solution of the IRLM in Sec.\ref{sec:BA}.

We now exploit the integrability of the boundary sine-Gordon model \eqref{eq:bsG}, which means that exact expressions for properties can be extracted.  We are interested in the dynamical energy scale generated by the (relevant) boundary term, which was calculated by Fendley, Ludwig and Saleur in Ref.~\onlinecite{FLSaleur}.  Taking Eq.~6.20 of this paper and making the association in notation $\lambda_{1}=t', \; \kappa = D, \; T_{B}=\pi T_{K}$, we arrive at the expression:
\begin{eqnarray}\label{exactresult}
t' D^{-d}=\frac{2^{d}}{4\pi} \Gamma(d) \left[ \pi d \,T_{K}  \frac{2\sqrt{\pi}(\lambda +1)\Gamma(\frac{1}{2}+\frac{1}{2\lambda})}{\Gamma(\frac{1}{2\lambda})} \right]^{1-d}
\end{eqnarray}
where $d=\beta^{2}/16\pi$ is the scaling dimension of the boundary operator given in \eqref{scaling} above, and $\lambda = \frac{1}{d}-1$.

Rearranging this, comparing with Eq.~\eqref{energyscale1}, and removing the explicit dependence of the scaling dimension $d$ in favour of the exponent $\alpha$ given in Eq.~\eqref{alpha}, we extract the pre-factor in the energy scale $T_K$:
\begin{eqnarray}\label{f_equation}
\tilde{f}(\alpha)=\frac{2^{\alpha}\pi^{\alpha - \frac{3}{2}}\Gamma(\frac{\alpha -1}{2})}{\Gamma(\frac{\alpha}{2})\big[\Gamma(\frac{\alpha - 1}{\alpha})\big]^{\alpha}}
\end{eqnarray}

There is one more step that must be taken before comparing to numerical results, which is to express the field theory high energy cutoff $D$ in terms of physical parameters.  We can do this by looking at the non-interacting case, where $\alpha=2$.  It is easy to show that Eqs.~\eqref{energyscale1}\eqref{f_equation} match the non-interacting result $T_{K}(\alpha = 2)=T_{0}=\pi\nu (t')^{2}$ if $D=4/\pi\nu$.  This allows us to finally express our energy scale $T_{K}$ as:
\begin{equation}\label{TKexact}
\nu T_{K}=f_1(\alpha)(\nu t')^{\alpha}
\end{equation}
where $f_1(\alpha)$ is given by
\begin{eqnarray}\label{Gammaexpression}
f_1(\alpha)=\frac{2^{2-\alpha}\pi^{2\alpha - 5/2}\Gamma(\frac{\alpha-1}{2})}{\Gamma(\frac{\alpha}{2})\big[\Gamma(\frac{\alpha-1}{\alpha})\big]^{\alpha}},
\end{eqnarray}
$\alpha$ is given by Eq.~\eqref{alpha} as always, and the $1$ in $f_1$ indicates $N=1$ lead.  A few useful values of this expression for future reference are $f_1(\alpha=2)=\pi$ and $f_1(\alpha=1)=4/\pi$.  The former is by design; the latter will be compared to the strong coupling expansion in Sec.~\ref{sec:strong}.

It is understood that Eq.~\eqref{TKexact} for the emergent energy scale $T_K$ is still just the leading term in a power series in the dimensionless hybridization $\nu t'$, however we maintain that this expression for the leading order term is exact.  We now demonstrate this by first comparing to numerics, and then by comparing to analytical results from a strong coupling expansion.

\subsection{Numerics; NRG and DMRG}
\label{sec:numerics}

\begin{figure}
\begin{center}
\includegraphics[width=3in]{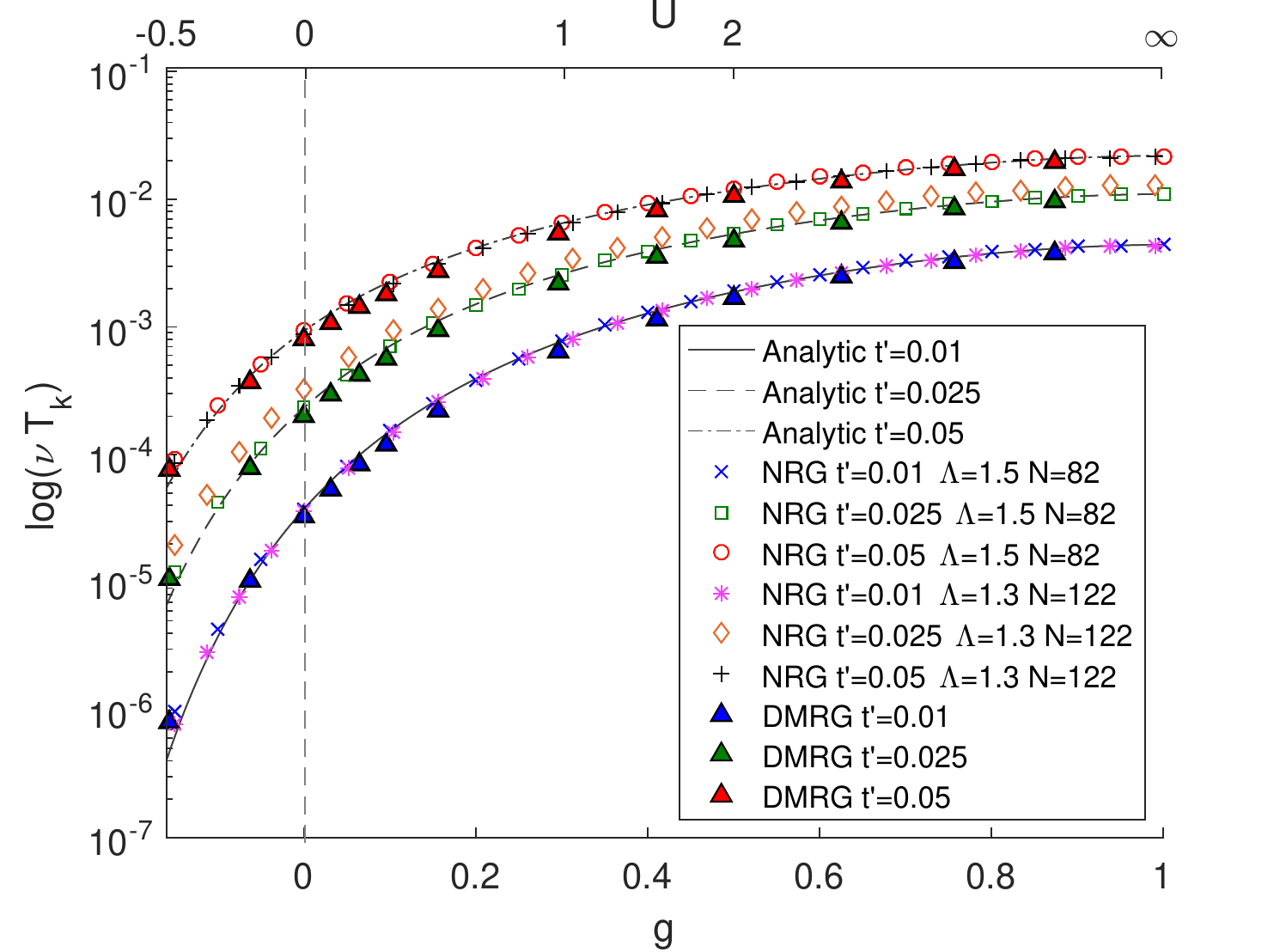}
\end{center}
\caption{
A comparison between numerical data from NRG and DMRG of the emergent energy scale given by (\ref{TKexact}) $\nu T_K$ as a function of interaction strength $g$ Eq.~\eqref{defg} and the analytic formula Eq.~\eqref{TKexact}. It is shown that the bigger the repulsive interaction $g$, the bigger the scale $T_{K}$, thus increasing the width of the dot occupation $n_{d}$. At $g=1$, for infinite repulsive interaction, the width reaches a finite maximum value. The correspoding values of $U$ with $g$ are written in the top axis.
}
\label{fig:Gamma_plot}
\end{figure}

There have been a number of previous numerical studies of the IRLM with the numerical renormalization group (NRG),\cite{Vinkler,Kiss1,Borda1,Borda2} and while good agreement is usually found for small interactions, there is usually a significant divergence for larger interactions.  Here, we present results from both NRG and density matrix renormalization group (DMRG) where we show that there is very good agreement with Eq.~\eqref{TKexact} with the exponent $\alpha$ given by Eq.~\eqref{alpha} and the pre-factor $f_{1}(\alpha)$ given by Eq.~\eqref{Gammaexpression} for all values of interaction $U$.

There are a few technical points about the numerical methods which are important for this model.  The first is that the Wilson leads of the NRG method mean that the density of states at low energy isn't completely constant.\cite{Peter_DMRG1}  To deal with this, for each individual value of $t'$, the appropriate value of $\nu$ is extracted from the non-interacting case, and then this is used through all further calculations with the same value of $t'$.  This issue can be alleviated by using tight binding leads in DMRG, however this leads to a different issue that there are very strong finite size effects in this case -- meaning that the DMRG uses a hybrid of tight-binding and Wilson leads.\cite{BohrSchmitteckertWoelfle:EPL2006,Schmitteckert:JPCS2010}  Details of both of these, along with other details of the numerical procedures are given in Appendix~\ref{sec:numapp}.

First, we extract the energy scale $T_K$ numerically -- this is done by numerically finding the slope of the line $n_d(\varepsilon_0)$ for each value of $t'$ and $g$; the results are plotted in Fig.~\ref{fig:Gamma_plot}.  It is seen that there is near-perfect agreement between the theoretical result \eqref{TKexact} and the numerics and we stress that there are no fitting parameters in this data aside from the NRG density of states determined from the non-interacting result.

\begin{figure}
\begin{center}
\includegraphics[width=3in]{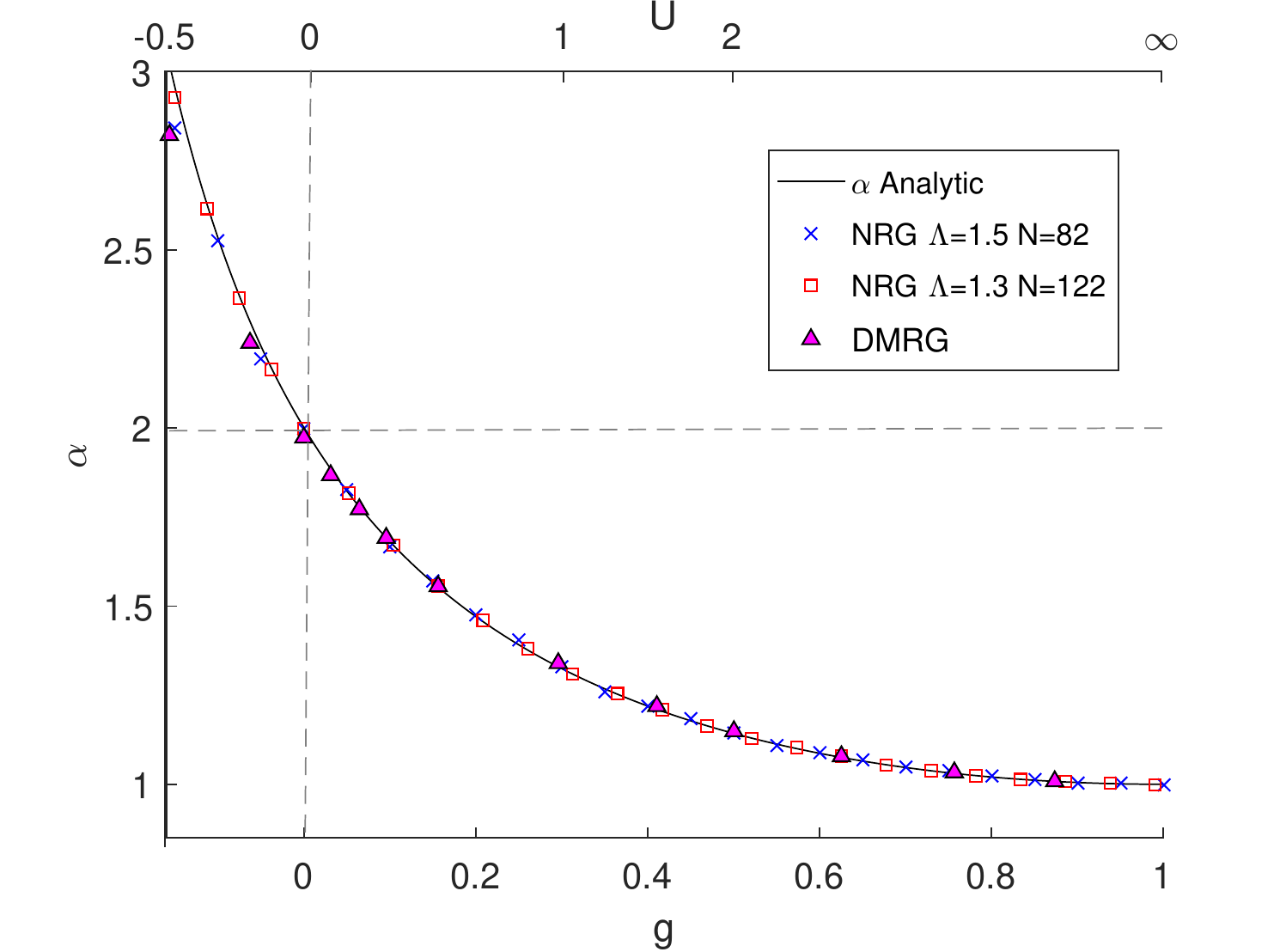}
\end{center}
\caption{DMRG/NRG data points calculated for the thermodynamic exponent $\alpha$ given by (\ref{alpha}) as a function of the interaction parameter $g$. The exponent was extracted by a set of $t'=0.01,0.025,0.05$. At $g=0$ the exponent gets the non-interacting limit value $\alpha=2$, whereas for $g>0$ decreases from this point to a finite value $\alpha=1$ at $U=+\infty$. In the region $g<0$, the exponent increases from the $g=0$ point, giving rise to a phase transition at $g=1-\sqrt{2}$ where $\alpha\to \infty$. The correspoding values of $U$ with $g$ are written in the top axis.}
\label{fig:Alpha_plot}
\end{figure}

\begin{figure}
\begin{center}
\includegraphics[width=3.5in]{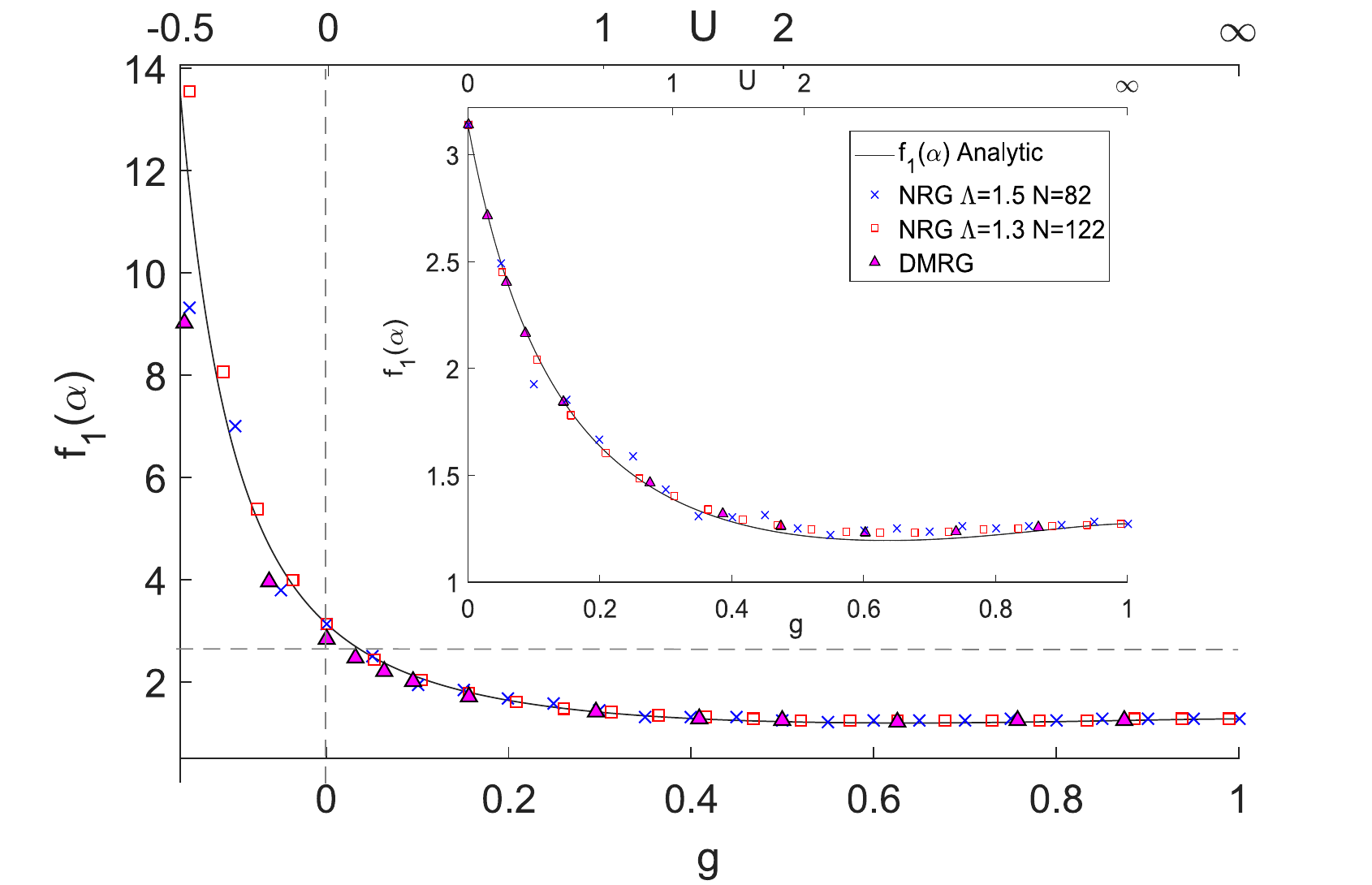}
\end{center}
\caption{NRG and DMRG data reproducing the prefactor $f_{1}(\alpha)$ given by equation (\ref{Gammaexpression}). The dashed lines have been included to mark the $g=0$ point for reference, which is the non-interacting limit. In the $g<0$ region, the prefactor increases its value, contributing to $T_{K}$ vanishing at the quantum phase transition point $g=1-\sqrt{2}$. \emph{Inset}: A zoom in the region $g\in[0,1]$, showing that $f_{1}(g)$ is nonmonotonic with increasing interaction.}
\label{fig:Prefactor_plot}
\end{figure}

We can also take the data for fixed $g$ and different values of $t'$ and fit to the form of Eq.~\eqref{TKexact} to numerically obtain the exponent $\alpha$ and the prefactor $f_1(\alpha)$ -- the results of which are plotted in Figs.~\ref{fig:Alpha_plot} and \ref{fig:Prefactor_plot} respectively.  Once again, the agreement is seen to be very good over the whole range of interactions.  We note that we can't go to strong negative values of the interaction as there is a quantum phase transition at $g\approx -0.21$ (corresponding in the lattice model to $U/t \approx -0.67$) where the exponent $\alpha$ diverges and one would expect a deviation from the result \eqref{TKexact}.  This is discussed further in Sec.~\ref{sec:QPT}.

\subsection{Strong coupling limit}
\label{sec:strong}

\begin{figure}
\begin{center}
\includegraphics[width=2.5in,clip=true]{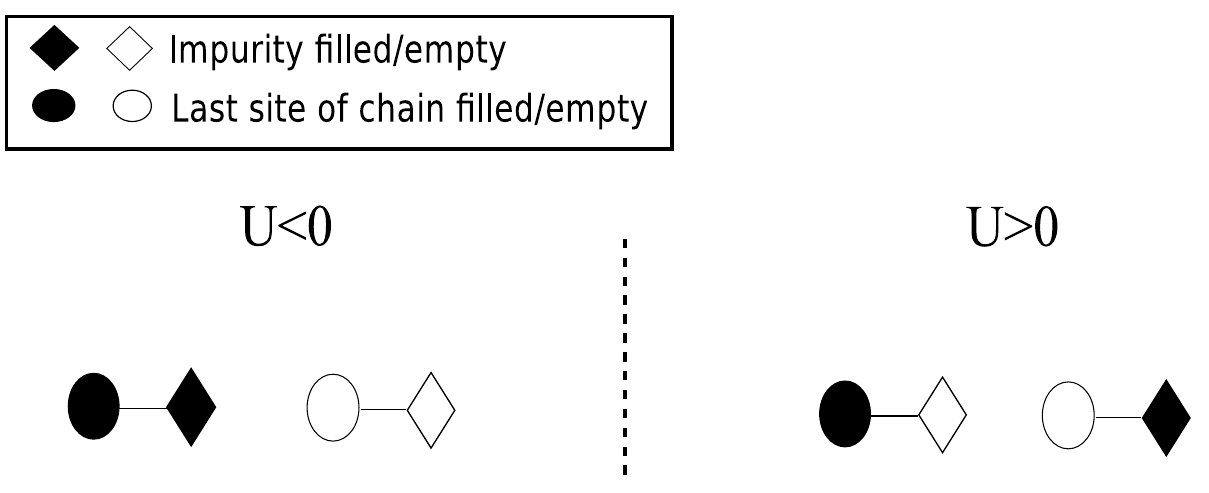}
\end{center}
\caption{Representation of the low energy subspace states in the strong coupling regime for $N=1$ in the repulsive and attractive cases.}
\label{fig:strongN1}
\end{figure}

While we have already shown that the expression Eq.~\eqref{TKexact} describes numerical data extremely well over the whole range of interactions, there is one more limit as $g\rightarrow 1$ (corresponding to $U\rightarrow \infty$) where we can show that this expression is exact.   In this limit, we can build up a low energy effective theory, by separating the impurity and the last site of the chain as a zeroth order approximation.  The strong repulsion between the two sites means at low energy, only one may be occupied, which is illustrated schematically in Fig.~\ref{fig:strongN1}.  In this low-energy subspace, any hopping between the end site of the chain and the rest of the chain is supressed, as this process necessarily takes the system out of the low-energy subspace.  This zeroth order approximation is therefore exact for $g=1$, and is the leading term in a $1/U$ expansion for finite interaction strength.  We will return to the $1/U$ corrections shortly.

The Hamiltonian describing the low-energy subspace of such a two-site system (after an unimportant constant shift in energy of $U/4$) is given by:
\begin{eqnarray}\label{hamiltonian_sc}
H_{\text{strong}}=\begin{pmatrix} \varepsilon_0 & t' \\ t' & 0 \end{pmatrix} = \varepsilon_0 \sigma_z + t' \sigma_x + \text{const.}
\end{eqnarray}
In this notation, the state $( 1 \; 0 )^T$ has the impurity level occupied (and hence feels the potential $\varepsilon_0$) and the last site on the lead unoccupied; while the state $(0\; 1)^T$ is the other way around.

The matrix is easily diagonalised to get the ground state wavefunction
\begin{eqnarray}
\psi_\text{GS} &=& \frac{1}{N} \begin{pmatrix} \varepsilon_{0}-\sqrt{\varepsilon_{0}^{2}+(2t')^{2}} \\  2t' \end{pmatrix} \nonumber\\
N &=& \sqrt{(2t')^{2}+\left(\varepsilon_{0}-\sqrt{\varepsilon_{0}^{2}+(2t')^{2}}\right)^{2}}
\end{eqnarray}
and hence the ground state occupation of the impurity level is given by
\begin{eqnarray}\label{occupation_x}
n_{d} =1-\frac{1}{1+(\varepsilon_{0}/2t'-\sqrt{1+(\varepsilon_{0}/2t')^{2}})^{2}}
\end{eqnarray}

By expanding around $\varepsilon_0 \sim 0$, we get:
\begin{eqnarray}
n_{d}\sim\frac{1}{2}-\frac{\varepsilon_0}{4t'}+ O((\varepsilon_0/t')^2)
\end{eqnarray}
and applying equation (4) gives the result:
\begin{eqnarray}
T_{K}(g= 1)=\frac{4 t'}{\pi}
\label{eq:TKg1}
\end{eqnarray}
which agrees perfectly with the general expressions \eqref{TKexact} and \eqref{Gammaexpression} in the limit $g\rightarrow 1$.  It is worth emphasising that like the numerics, Eq.~\eqref{eq:TKg1} was calculated on a lattice, while the general expression came from field theory -- and yet they match.

We can go further than this by now making a perturbation expansion in $1/U$ around this $U \rightarrow \infty$ limit.   
Using a Schrieffer-Wolff transformation\cite{SW} (or equivalently second-order perturbation theory), we find that to leading to an effective hamiltonian describing the low energy physics in the strong coupling regime:
\begin{eqnarray}
H_{\text{eff}}=t' \sigma^{x}+ \varepsilon_0 \sigma^{z}+H_0+\frac{4t^{2}}{U} \sigma^{z}\bigg(c_{0}^{\dagger}c_{0}-\frac{1}{2}\bigg)
\label{eq:HafterSW}
\end{eqnarray}
The first two terms are the same as in Eq.~\eqref{hamiltonian_sc} above; the next term is the free Hamiltonian of the lead (minus the final site, but in the infinite size limit this makes no difference), and the final term is the leading correction from virtual hopping between the lead and the dot and back again; $c_0$ now refers to the last site of the new lead, which is the last-but-one site of the original lead.

The Hamiltonian is similar in form to that of the IRLM, except that the $t' \sigma^x$ term is not coupled to the lead.  The interaction between the lead and the (enlarged effective) impurity site however can be removed by applying exactly the same bosonization and unitary transformation steps as in Sec.\ref{sec:bosonization}; the resulting Hamiltonian is
\begin{multline}\label{strongcouplingH}
H=\frac{1}{2}\int_{-\infty}^{+\infty}dx(\partial_{x}\phi(x))^{2} + \varepsilon_0 \sigma_{z} \\
+ t' \bigg(S^{+}e^{-i\sqrt{8\pi d'}\phi(0)}  +\text{h.c}\bigg)
\end{multline}
where the scaling of vertex operators is
\begin{equation}
d'=\frac{1}{2}\left(\frac{4t}{\pi U}\right)^2.
\end{equation}
As before, this gives an emergent energy scale $T_K \sim t'^\alpha$ where $\alpha=1/(1-d') \approx 1+ d'$ for large $U/t$.  Expanding Eq.~\eqref{phase} to get the phase shift $\delta \approx \pi/2 - 2t/U$, giving $g \approx 1 - 4t/\pi U$ and substituting into Eq~\eqref{alpha} gives exactly the same expression -- hence we have analytically proven that the exponent $\alpha$ for strong interactions matches the general expression to order $(t/U)^2$.

We can also look at the strong coupling on the attractive side $U\rightarrow -\infty$, where we again single off the last site of the lead along with the dot and the low-energy subspace is now either both occupied or both empty, see Fig.~\ref{fig:strongN1}.  In this case, the hybridization $t'$ doesn't enter the low-energy Hamiltonian -- the ground state is both occupied for $\varepsilon_0 <0$, and both occupied otherwise, with a degeneracy (resonance) at $\varepsilon_{0}=0$.  The resonance width $T_K$ is thus strictly $0$.  As with the repulsive case, one can also perturb around this point.  In this case, to induce a transition from one state to the other, one must hop two electrons to the lead, $(t^2/U) \psi^{\dagger}(j=-1)\psi^{\dagger}(j=-1)\sigma^{-}$ or vice versa.  This operator has scaling dimension $d=2$ and is thus strongly irrelevant -- and hence the resonance width remains $T_K=0$.  This is all consistent with a quantum phase transition for some critical attractive interaction which is far away from the strong coupling limit -- more will be said about this in Sec.~\ref{sec:QPT}.

To summarise so far, we have derived an exact expression, Eq.~\eqref{TKexact} for the resonance width of the $N=1$ IRLM.  We have shown that this matches numerical results as well as agreeing with the leading terms of the strong coupling expansion.  We now go beyond the width and look at the resonance line shape.

\section{Bethe ansatz for the occupation of the dot}
\label{sec:BA}

\begin{figure*}
\includegraphics[width=0.9\textwidth]{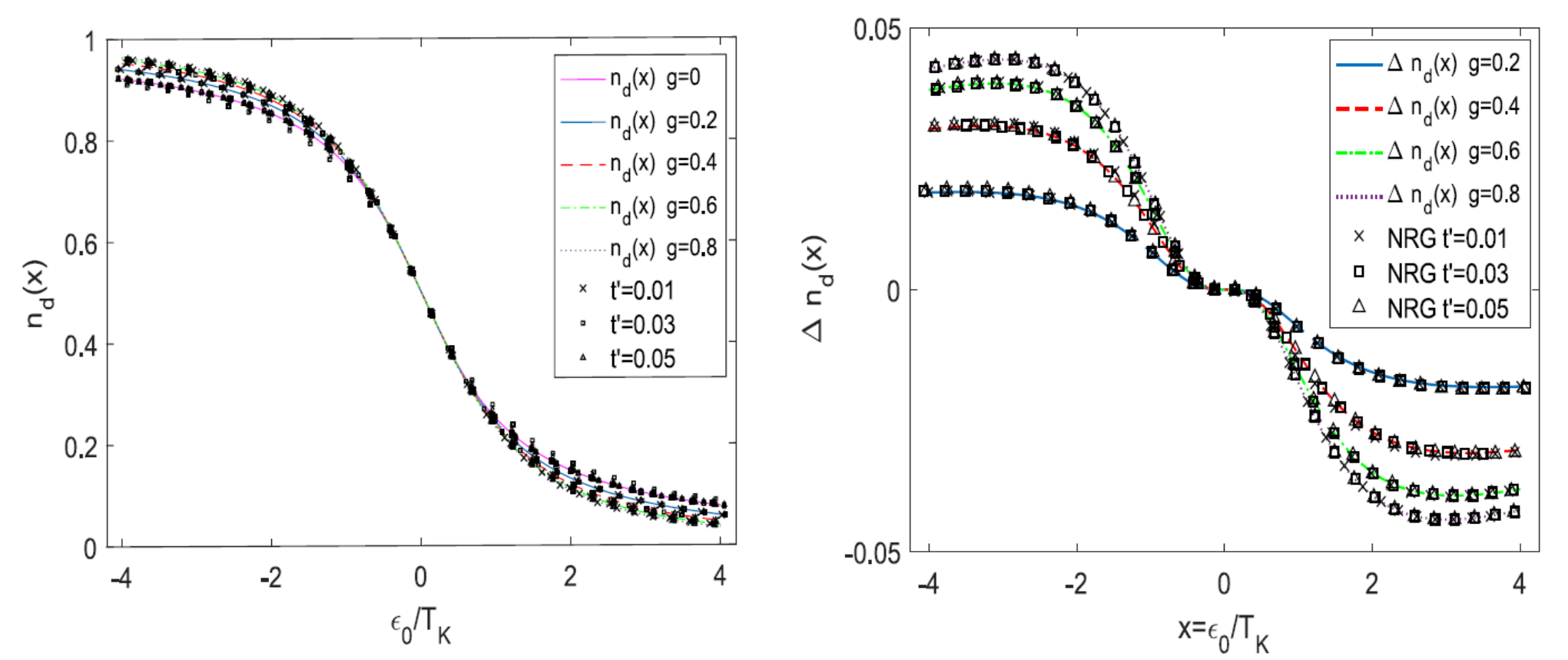}
\caption{Left: NRG simulations (data points) for the occupancy of the dot $n_{d}$ as a function of the scaled variable $x$, compared with the analytical solution (lines), Eq.~\eqref{nd_BA_equation}. The difference between curves for different interaction strength is not well seen on this scale. Right: the same data points, but now the non-interacting occupation has been subtracted ($\Delta n_{d}(x)=n_{d}(x,g\neq 0)-n_{d}(x,g=0)$) to emphasize the small changes in shape as interaction strength is varied. The curves show universal scaling with the variable $x$ for different values of hybridization $t'$, and emphasizes the importance of the width $T_{0}$ in the scaling.}\label{fig:nd}
\end{figure*}

The IRLM (in the continuous lead version) is integrable and has been solved by the Bethe ansatz.\cite{FilyovW,Ponomarenko,Rylands}  In particular, as well as obtaining the width of the resonance, the exact line shape $n_d(\varepsilon_0)$ can be calculated.  Taking either Eq.~17 of Ref.~\onlinecite{Ponomarenko} or Eq.~46 of Ref.~\onlinecite{Rylands} and making a switch to our notation (in Ref.~\onlinecite{Rylands}, this amounts to equating our scaling dimension $\alpha$ to their $\Delta$ through the relationship $\alpha=\pi/\Delta$ which can be seen in Eq.~45 of their paper; in Ref.~\onlinecite{Ponomarenko} the notation is harder to penetrate but it gives the same answer) gives
 \begin{multline}\label{nd_BA_equation}
 n_{d}(\varepsilon_0) = \frac{1}{2}-\frac{1}{\sqrt{\pi}}\sum_{n=0}^{\infty}\frac{(-1)^{n}}{n!}\frac{\Gamma\big(1+\frac{\alpha}{2}(2n+1)\big)}{\Gamma\big(1+\frac{\alpha-1}{2}(2n+1)\big)} \\
 \times  \frac{1}{2n+1}  \left( \frac{\varepsilon_0}{T_B} \right)^{2n+1}.
 \end{multline}
This sum converges for $\varepsilon_{0}/T_B$ less than some critical value; for completeness the expansion at large values of this parameter is given in Appendix \ref{sec:appBA}. It is worth pointing out that the same series is found in the Anisotropic Kondo Model from the exact solution by Bethe ansatz\cite{TsvelickWiegmann,AKM_1,AKM_2,Fendley_AKM} for the impurity magnetization, where different notations are used. This expression is written in terms of a boundary energy scale $T_B$ which can be easily related to our $T_K$ by looking at the $n=0$ term and using the definition of $T_K$, \eqref{Gammades}.  This gives:
\begin{eqnarray}\label{boundary_temp_relation}
T_{B}=\sqrt{\pi}\frac{\Gamma(1+\alpha/2)}{\Gamma(1/2 + \alpha/2)}\, T_{K}
\end{eqnarray}

In Appendix \ref{sec:appBA}, we show that the general expression \eqref{nd_BA_equation} can be resummed in two cases: (i) the non-interacting case $\alpha=2$, where it reproduces the known result \eqref{nd-U0}; and (ii) in the $U\rightarrow \infty$ case $\alpha=1$, where it gives our previously derived strong coupling result \eqref{occupation_x}.

We can also compare it to numerical results which is shown in Fig.~\ref{fig:nd}.  Firstly, it is worth pointing out that when the occupation is studied as a function of the scaling variable $\varepsilon_0/T_K$ with $T_K$ defined via the inverse slope at $\varepsilon_0=0$, Eq.~\eqref{Gammades}, the function $n_d(\varepsilon/T_K)$ looks very similar for all values of interaction.  However we can also focus on the differences by subtracting off the non-interacting expression, where near-perfect agreement between numerics and the analytic result \eqref{nd_BA_equation} is seen.

\subsection{A comment on the Bethe-ansatz solution}
\label{sec:BAcomment}

\begin{table}
\begin{center}
\begin{tabular}{c|c|c}
 Author & Non-interacting leads & LL leads \\ \hline\hline
 Filyov and Wiegmann\cite{FilyovW} & $\frac{1}{2} - \frac{U}{2\pi} \approx \frac{1}{2} - \frac{g}{2}$ & --- \\
 Ponomarenko\cite{Ponomarenko} & $\frac{1}{2} - g$ & --- \\
  Rylands and Andrei\cite{Rylands} & $\frac{1}{2} - \frac{g}{2}$ & $\frac{1}{2K} - \frac{g}{2}$ \\
 This work & $\frac{(1-g)^2}{2} = \frac{1}{2} - g + \frac{g^2}{2}$ &  $\frac{(1-g)^2}{2K}$
\end{tabular}
\end{center}
\caption{A comparison of different works calculating the scaling dimensions $d$ giving rise to the exponent $\alpha=1/(1-d)$ as defined in Eq.(\ref{alpha}).  We have converted notation in other papers into our parameterisation of the interaction strength $g=\frac{2}{\pi}\arctan (U/2)$ which is approximately $U/\pi$ for small $U$, which for convenience is written as a dimensionless variable (equivalent to taking the hopping $t=1$ in our paramaterisation of the model).  The first column is for a non-interacting leads as discussed in the majority of this work; the second column is for coupling to Luttinger liquid (LL) leads.  The first three works get the exponent from the Bethe ansatz while this work derives it from bosonization.}\label{tab:exponents}
\end{table}

From the excellent agreement of Eq.~\eqref{nd_BA_equation} to numerics as well as non-trivial analytic limits may be interpreted as confirmation that the Bethe-ansatz calculations for properties of the one-channel IRLM are correct, there is actually a big caveat here.  We specifically wrote the dot occupation \eqref{nd_BA_equation} as a function of the exponent $\alpha$ rather than as a function of interaction strength $g$.  This is partially because the expressions are a bit shorter written in this way, but most importantly, the Bethe ansatz \textit{does not} get the relationship between $g$ and $\alpha$ correct, as summarised in Table~\ref{tab:exponents}.

The table extracts the relationship between the interaction and the exponent from three previous Bethe ansatz studies and compares them to the results in this work.  The oldest study by Filyov and Wiegmann\cite{FilyovW} doesn't introduce the phase shift, but this can be accounted for by the small $U$ expansion $g \approx U/\pi$.  It is seen that (i) none of the Bethe ansatz studies find the $g^2$ term in the exponent; and (ii) two of them do not  find the correct first order term.  It is worth pointing out however that $d=1/2-g/2$ however does give the correct strong coupling limit of $d=0$ at $g=1$, unlike $d=1/2-g$ which is correct for small interactions but fails to reproduce the strong coupling limit.  We also note that while this paper concentrates on non-interacting leads, the recent Bethe-Ansatz results of Rylands and Andrei\cite{Rylands} are for coupling to a Luttinger liquid  (LL), but the way the Luttinger parameter $K$ enters the expression for the exponent also does not agree with the result from bosonization, which is derive in Appendix~\ref{sec:appLL}. Testing the prediction for coupling to a LL would be numerically challenging and is to date an open question, although previously reported results have opened the path towards this generalisation by claiming universality of thermodynamics to hold in the presence of bulk interactions.\cite{Berkovits2}

However, the most important detail we wish to focus on is the failure of the Bethe-ansatz to reproduce the $g^2$ term in the exponent.  This is particularly important as this is the only term that changes when more leads are added $N>1$ as we will see in Sec.~\ref{sec:Nbiggerthan1}.  We therefore defer further discussion of this point to Sec.~\ref{sec:BAcomment2}.

\section{Aside: The Toulouse point of the Kondo model}
\label{sec:Kondo}

Let us now comment briefly on the relationship between the IRLM and the Anisotropic Kondo Model (AKM), from which the IRLM was first introduced.\cite{WiegmannFinkelshtein} In particular, we will focus on the point where the AKM maps onto the Resonant Level Model (RLM), which is the $U=0$ point of the IRLM. The anisotropic version of the Kondo hamiltonian has an XXZ coupling between the impurity spin and the conduction electrons; the simplest version of it reads\cite{Gogolin}
\begin{eqnarray}\label{AKMham}
H=H_{0}+ \frac{J_{xy}}{2}\big(s^{+}S^{-}(0) + \text{h.c}\big) +J_{z}s^{z}S^{z}(0)
\end{eqnarray}
where $H_{0}$ represents the (spinful) conduction electrons, $s^{\gamma}$ ($\gamma=+,-,z$) represents the impurity spin, and $S^{\gamma}$ represent the fermions density coupled to the impurity at the origin:
\begin{eqnarray}
S^{+}(x=0)&=&:\psi^{\dagger}_{\uparrow}(0)\psi_{\downarrow}(0): \nonumber\\
 S^{z}(x=0)&=&\frac{1}{2}:\big(\psi^{\dagger}_{\uparrow}(0)\psi_{\uparrow}(0)-\psi^{\dagger}_{\downarrow}(0)\psi_{\downarrow}(0)\big): 
\end{eqnarray}
Note that these fermionic operators describe only right moving fields, after proper unfolding in the system, in analogy with the mapping of eq.~(\ref{H_single_channel}). Under a conveniently chosen unitary transformation on (\ref{AKMham}), the AKM maps onto a spinless version of a resonant level hybridizing with the rest of the system, which is just the IRLM at $U=0$. The particular value of the coupling $J_{z}$ for this to happen is what determines the Toulouse point. This particular value of the coupling is of interest since it makes the model non-interacting. We proceed now to detail its calculation, since different values of the Toulouse point have been reported in the literature.\cite{Gogolin,Fabrizio}

To begin with, let us start by the case where $J_{xy}=0$. At this point, the model is equivalent to that of a scattered particle in a delta barrier at the origin, with each projection $s^{z}=\pm 1/2$ of the hamiltonian constituting a central potential scattering problem. The potential felt by conduction electrons reaching the boundary depends on the impurity spin orientation $\pm \frac{J_{z}}{2}s^{z}$. This is a single particle problem that can be solved exactly, in the same line as it was described in II.B. In the limit $J_{xy}=0$ treated here, all conduction electrons experience a phase shift on the wavefunction equal to:
\begin{eqnarray}\label{delta_AKM}
\delta=\arctan\bigg(\frac{J_{z}\pi\nu}{4}\bigg) && g=\frac{2\delta}{\pi}
\end{eqnarray}
where we have defined again the coupling $g\in [-1,1]$, this time with $\delta$ given by eq.~(\ref{delta_AKM}). Note that respect to the IRLM eq.~(\ref{phase}), this phase shift includes a factor of $1/2$ as a consequence of spin. The same procedure used in III.A is applied here to the AKM. Our first step is to bosonize the above hamiltonian by using the fermion/boson correspondence given by eq.~(\ref{bosonization_relation}). One just has to keep in mind that now spin is present, and therefore two relations of the type of (\ref{bosonization_relation}) are needed, one for each spin species.

In its bosonized form, hamiltonian (\ref{AKMham}) reads:
\begin{eqnarray}
H&=&H_{0}+\frac{J_{xy}\delta(x)\eta_{\uparrow}\eta_{\downarrow}}{4\pi}\big(s^{+}e^{i\sqrt{4\pi}(\phi_{\uparrow}(x)-\phi_{\downarrow}(x))}+\text{h.c}\big)\nonumber\\
&+&\frac{J_{z}\delta(x)}{2\sqrt{\pi}}(\partial_{x}\phi_{\uparrow}(x)-\partial_x\phi_{\downarrow}(x))s^{z}
\end{eqnarray}
where the non-interacting part $H_{0}$ is the sum of two non-interacting (bulk) baths of fermions, each carrying different spin projection. We note that written in this way, one can choose a more conveniently basis of the operators to work with, since only the antisymmetric combination of the fields is coupled to the impurity\cite{Gogolin}:
\begin{eqnarray}
\phi(x)=\frac{1}{\sqrt{2}}\big(\phi_{\uparrow}(x)-\phi_{\downarrow}(x)\big)
\end{eqnarray}
We will not bother about the Klein factors here and will choose the representation $\eta_{\uparrow}\eta_{\downarrow}=+1$. Then the bosonized version of the hamiltonian in the rotated system only contains the antisymmetric field $\phi$ (and a totally decoupled part that stands from the symmetric linear combination of $\phi_{\uparrow}(x)$ and $\phi_{\downarrow}(x)$):
\begin{eqnarray}\label{AKM_bos}
H=H_{0}[\phi] + \frac{J_{xy}}{4\pi}\big(s^{+}e^{i\sqrt{8\pi}\phi(0)}+\text{h.c}\big)\nonumber\\
+\frac{\sqrt{2}J_{z}}{2\sqrt{\pi}}s^{z}\partial_{x}\phi(0)
\end{eqnarray}
\begin{figure}
\begin{center}
\includegraphics[width=3in,clip=true]{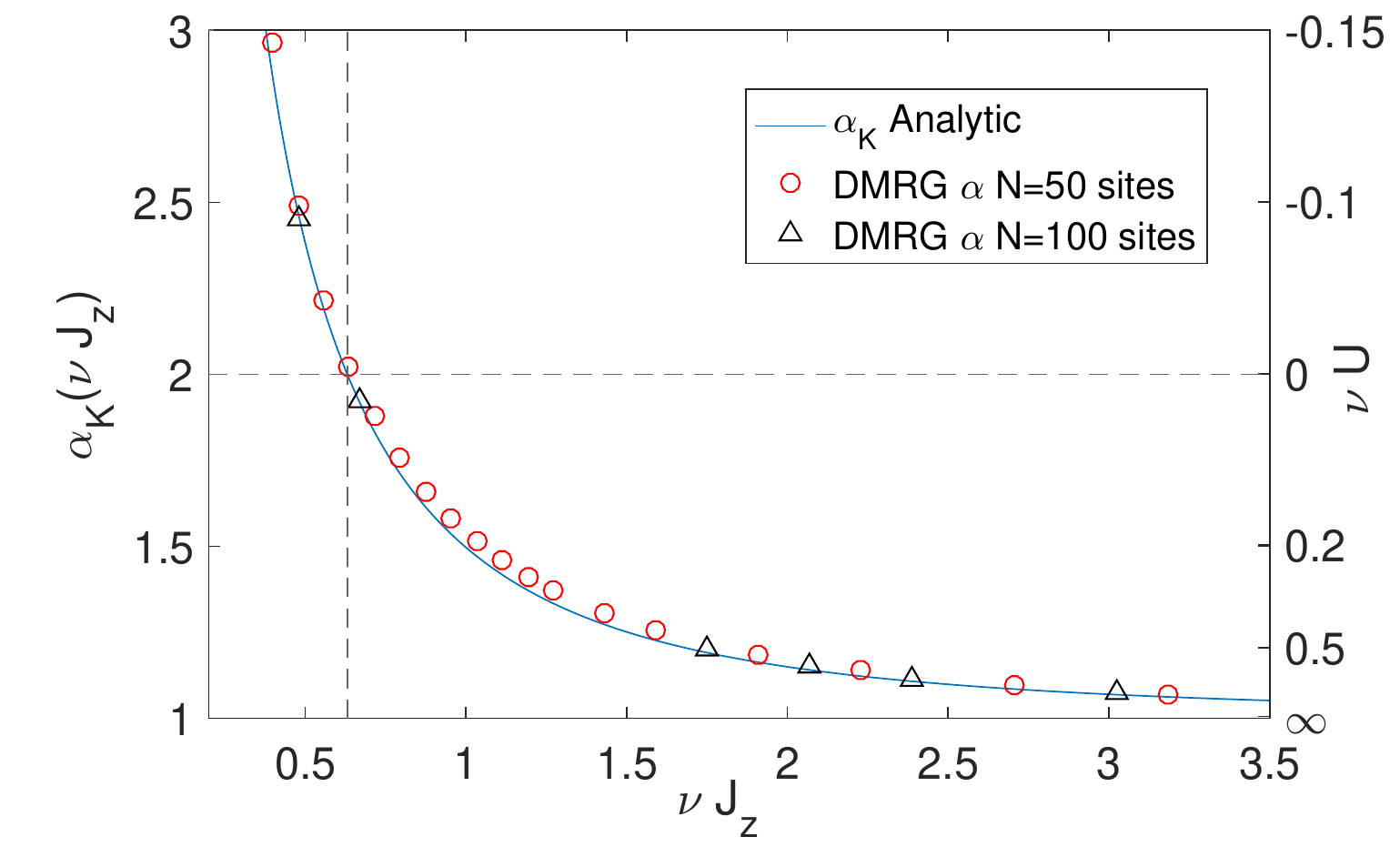}
\end{center}
\caption{Numerical results from DMRG for the exponent $\alpha_{K}$ in the Anisotropic Kondo Model (AKM). The Toulouse point has been marked with dashed lines, and it is given by equation (\ref{TP_value}). The usual value found in the literature\cite{Gogolin} is situated at the point $\nu J_{z}=2\pi\sqrt{2}(\sqrt{2}-1)\sim 3.68$, where here $\nu J_{z}\approx 0.63$}
\label{fig:Toulouse_point_figure}
\end{figure}

 The microscopic parameter of the theory $J_{z}$ is now substituted by the equivalent single-particle problem, where $J_{z}/2$ enters straight into the scattering phase shift of conduction electrons. We make use of the \emph{phase shift substitution} employed previously in section III.A, and work with the convenient interaction parameter $g$ as defined in eq.~(\ref{delta_AKM}). This allows us to work with $J_{xy}\neq 0 $ so long as the condition $\nu J_{xy}\ll 1$ is preserved. This leaves the hamiltonian in the following form:
\begin{eqnarray}\label{AKM_2}
H=H_{0}[\phi] + \frac{J_{xy}}{4\pi}\big(s^{+}e^{i\sqrt{8\pi}\phi(0)}+\text{h.c}\big)\nonumber\\
+\sqrt{2\pi}gs^{z}\partial_{x}\phi(0)
\end{eqnarray}

where we have used the phase shift substitution $J_{z}/2\to 2\delta$. This is because it is $J_{z}/2$ (and not $J_{z}$) the actual amplitude of the potential barrier seen by conduction electrons (see hamiltonian (\ref{AKMham})). 

One proceeds now as before and applies a unitary transformation to (\ref{AKM_2}) to get rid of the part proportional to $g$ by changing the scaling dimension of the vertex operators. The applied unitary transformation is similar in form to (\ref{unitary_transf}):
\begin{eqnarray}\label{unitary_AKM}
\mathcal{U}=e^{i\sqrt{8\pi}gs^{z}\phi(0)}
\end{eqnarray}
The above unitary transformation is then applied to the hamiltonian, and the coupling $g$ is absorbed into the scaling dimension of the vertex operators, thus getting rid of the part proportional to $\partial_{x}\phi(x)$. After applying the unitary transformation (\ref{unitary_AKM}) into hamiltonian (\ref{AKM_2}) we obtain:
\begin{eqnarray}
H=H_{0}+\frac{J_{xy}}{4\pi}(s^{+}e^{i\sqrt{8\pi}(1-g)\phi(0)}+\text{h.c})
\end{eqnarray}
The scaling dimension of the vertex operator is now $d=(1-g)^{2}$. Thus the associated exponent of the theory (in analogy with $\alpha$ in equation (\ref{alpha})) reads now:
\begin{eqnarray}\label{Alpha_AKM}
\alpha_{K}=\frac{1}{1-d}=\frac{1}{2g-g^{2}}
\end{eqnarray}

The numerical extraction of $\alpha_{K}$ by DMRG is represented in Fig.~\ref{fig:Toulouse_point_figure}. In order to end up with fermions, the scaling dimension of vertex operators has to be $d=\frac{1}{2}$. This means:
\begin{eqnarray}
d=\frac{\beta^{2}}{8\pi}=(1-g)^2=\frac{1}{2}
\end{eqnarray}

 This happens for a specific value of the coupling parameter $g$ ($g\in[-1,1]$), which will determine the value of the Toulouse point:
\begin{eqnarray}\label{TP_value}
g&=&1-\frac{1}{\sqrt{2}}\nonumber\\
\nu J_{z} &=&\frac{4}{\pi}\tan\bigg(\frac{\pi g}{2}\bigg)\approx 0.63
\end{eqnarray}

This concrete value of $\nu J_{z}$ maps the anisotropic Kondo model into a non-interacting resonant level model, that is, to hamiltonian (\ref{H_single_channel}) at $U=0$. In Fig.~\ref{fig:Toulouse_point_figure} we identify this point by the light dashed lines, while DMRG results on $\alpha_{K}$ prove to be in good agreement with eq.~(\ref{Alpha_AKM}). The Toulouse point differs numerically from the value given in previous works.\cite{Gogolin,Fabrizio} The susbtitution of the interaction coupling by the  scattering phase shift is an essential step to derive that result.\\

\section{THE MULTICHANNEL IRLM $N>1$}
\label{sec:Nbiggerthan1}

Let us now turn to the multichannel version of the IRLM, described by Hamiltonian \ref{Hlattice} when $N>1$.  This includes the case $N=2$ which is particularly important for the study of transport properties,\cite{Andrei,Boulat1,Boulat2,Sampaper,Sampaper2,Sampaper3,Vinkler,Borda1,vonDelftIRLM2018} but it is also instructive to study the model for generic $N$.\cite{Borda2,Kiss2}

By making a Fourier transform with respect to chain index
\begin{equation}
c_{n,\kappa} = \frac{1}{\sqrt{N}}\sum_{\gamma=1}^N e^{i \sqrt{2\pi/N} \kappa \gamma} c_{n,\gamma}
\end{equation}
the Hamiltonian becomes
 \begin{eqnarray}\label{HlatticeN}
 H=H_{0}+\varepsilon_{0}d^{\dagger}d + \sqrt{N} t' \bigg( d^{\dagger}c_{0,\kappa=0} +{\text{h.c}}\bigg)\nonumber\\ 
 + U\sum_{\kappa=0}^{N-1}\bigg(d^{\dagger}d -\frac{1}{2}\bigg)\bigg(c_{0,\kappa}^{\dagger}c_{0,\kappa}-\frac{1}{2}\bigg) 
 \end{eqnarray}
with
 \begin{eqnarray}
 H_{0}=  -t   \sum_{\kappa=0}^{N-1}  \sum_{i=0}^{+\infty}  c_{i+1,\kappa}^{\dagger}c_{i,\kappa}+\text{h.c},
 \label{eq:H0latticeN}
 \end{eqnarray}
 in other words, the impurity is only hybridized with the $\kappa=0$ mode, but the interaction still couples to all of the other channels.
 
 In the non-interacting case, this therefore maps back exactly onto the single-channel case, with the only difference that the hybridization $t'$ is scaled by $\sqrt{N}$.  Hence the trivial generalisation of Eq.~\eqref{non_interacting_T} to the multichannel case is
 \begin{equation}
 T_0 = N\, \pi\nu (t')^2.
 \end{equation}
 We therefore extend our parameterisation \eqref{TKexact} of the resonance width $T_K$ to the multichannel case as
 \begin{equation}\label{TKexactN}
\nu T_{K}=f_N(g)(\sqrt{N}\nu t')^{\alpha_N(g)}
\end{equation}
so that in the non-interacting limit $f_N(g=0)=\pi$ and $\alpha_N(g=0)=2$, independent of the number of channels.  In this section, we will look at the behavior of the exponent $\alpha_N(g)$ and the pre-factor $f_N(g)$ as a function of interaction strength $g$ for various values of $N$.

\subsection{Bosonization}
\label{sec:bosonizationN}

Following Sec.~\ref{sec:bosonization} we now take the Hamiltonian of the multichannel IRLM which for convenience we use the form \eqref{HlatticeN} above, linearize the spectrum, unfold the fields, and bosonize it according to relation \eqref{bosonization_relation}.   Ignoring the Klein factors which are non-dynamic as in the single channel case gives $H=H_0 + H_U + H_{t'}$ where
\begin{eqnarray}
H_{0} &=& \frac{1}{2} \sum_{\kappa=0}^{N-1} \int_{-\infty}^{+\infty} dx(\partial_{x}\phi_\kappa(x))^{2} \nonumber \\
H_U &=& \sqrt{\pi} g \sum_{\kappa=0}^{N-1} \partial_{x}\phi_\kappa(0)S^{z} \nonumber \\
H_{t'} &=& \sqrt{\frac{N}{2\pi}} t' \big(e^{-i\sqrt{4\pi}\phi_0(0)}S^{-}+\text{h.c}\big)+\varepsilon_0 S^z.
\end{eqnarray}
Here, we have already used the `phase-shift substitution' using $g=2\delta/\pi$ instead of $U$ to match the $t'=0$ case, and the only difference with the single-channel case is the sum over channels.  

\begin{figure}
\begin{center}
\includegraphics[width=3in,clip=true]{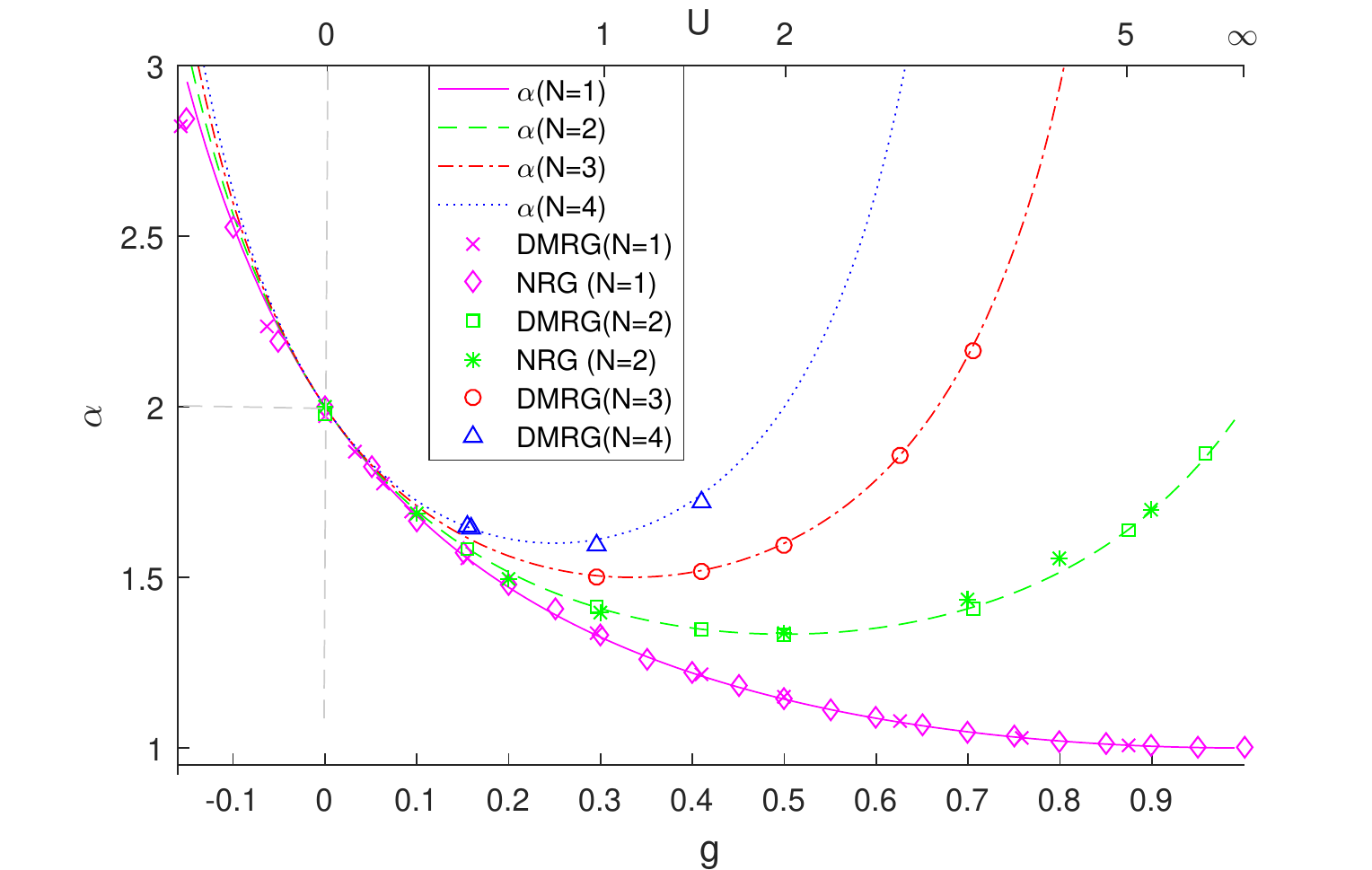}
\end{center}
\caption{Representation of the scaling exponent $\alpha$ for the thermodynamic energy scale as a function of both interacting parameters $g$ and $U$, obtained numerically by the Density Matrix Renormalization Group and NRG, showing data points matching eq.\eqref{alphaN} The non-interacting point in the model $U=0$ is pointed by the light dashed lines as a guide to the eye.}\label{fig:alphaN}
\end{figure}

\begin{figure}
\begin{center}
\includegraphics[width=3in,clip=true]{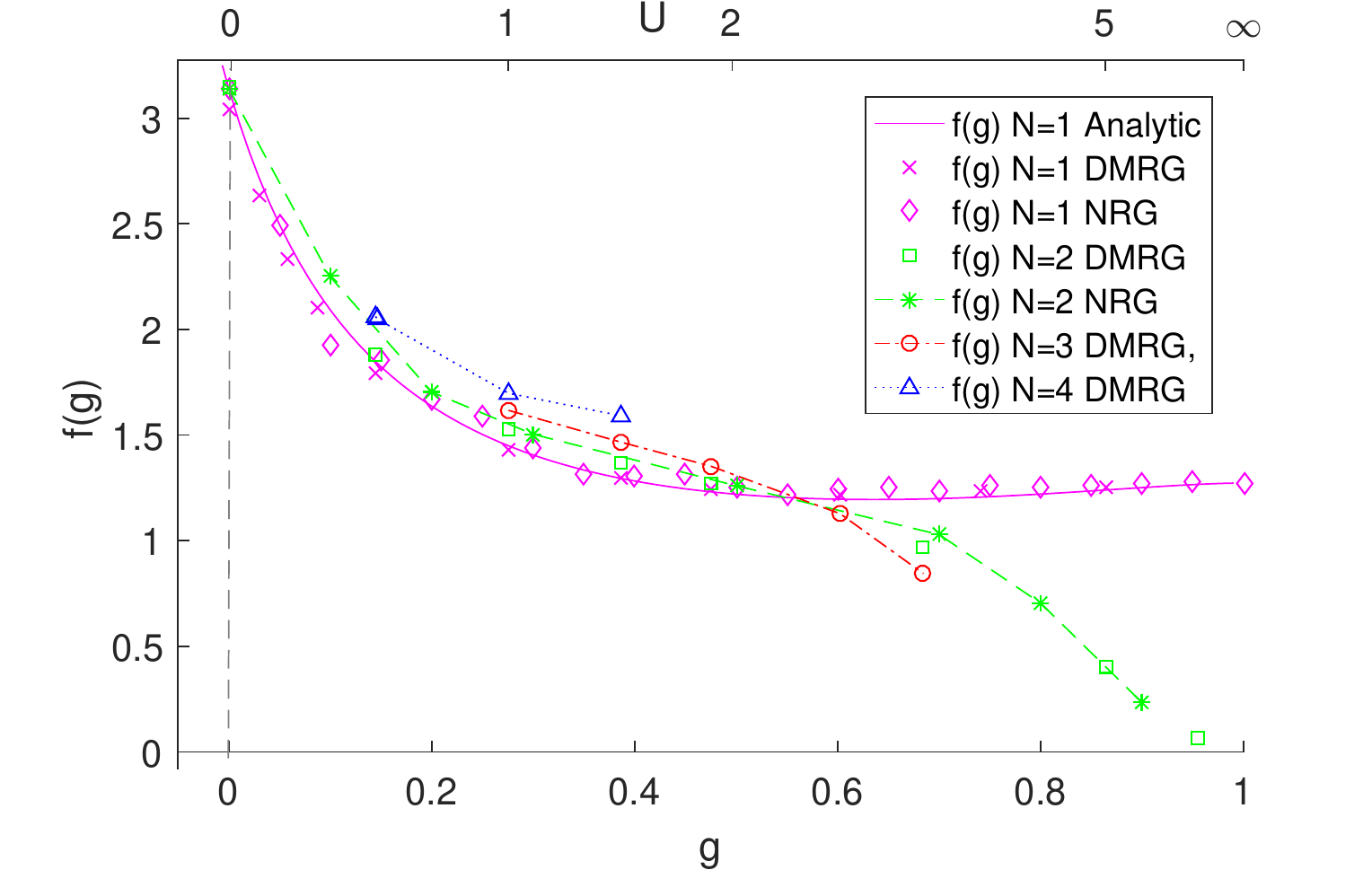}
\end{center}
\caption{The prefactor $f_{N}(g)$ for the multi-channel IRLM determined by DMRG($N=1,2,3,4$) and NRG($N=1,2$). Data points correspond to the same fitting procedure used to determine $\alpha$ in Fig.~\ref{fig:alphaN}}.
\label{fig:fN}
\end{figure}

Continuing to follow Sec.~\ref{sec:bosonization}, we eliminate the $H_U$ term by the unitary transformation
\begin{eqnarray}
\mathcal{U}=e^{i\sqrt{4\pi}g S^{z}\sum_{\kappa=0}^{N-1}\phi_{\kappa}(0)}
\end{eqnarray}
which doesn't affect $H_0$, but does modify the hybridization term:
\begin{multline}
H_{t'} \rightarrow \sqrt{\frac{N}{2\pi}} t' \left[ e^{-i\sqrt{4\pi}(1-g)\phi_0(0)} \left(\prod_{\kappa=1}^{N-1} e^{i\sqrt{4\pi} g \phi_\kappa(0) }\right) S^{-} \right.\\
\left. +\text{h.c} \right] +\varepsilon_0 S^z.
\label{eq:hybNtrans}
\end{multline}
The scaling dimension of the vertex operator can now be read off:
\begin{eqnarray}\label{dual_eq}
d&=&\frac{1}{2} \left( (1-g)^2 + (N-1) g^2 \right) \nonumber \\
&=&\frac{1}{2}\left(1-2g + Ng^{2}\right)
\end{eqnarray}
which gives the exponent $\alpha=1/(1-d)$ in the general relationship \eqref{TKexactN} to be
\begin{equation}\label{alphaN}
\alpha_N(g) = \frac{2}{1+2g-Ng^2}
\end{equation}
This is plotted for $N=1,2,3,4$ in Fig.~\ref{fig:alphaN} along with numerical data from DMRG obtained using the same procedure previously described in Sec~\ref{sec:numerics}. We have also extended the NRG procedure used in section III.C to include NRG data for the $N=2$ channel case. It can be seen that the agreement is very good across the entire range of interaction strength $g$.  While we are not aware of an analytic prediction for the pre-factor $f_N(g)$ for $N>1$, this can also be extracted numerically and is plotted in Fig.~\ref{fig:fN}.  We will come back to this shortly.

Let us now make a few observations about the properties of the multichannel IRLM.  Firstly, the scaling exponent $d$ in Eq.~\eqref{dual_eq} may also be written
\begin{equation}
d = \frac{1}{2} - \frac{1}{2N} + \frac{1}{2N} (1-Ng)^2 .
\end{equation}
This illustrates a few different things:
\begin{enumerate}
\item At $g=1/N$, the scaling dimension $d$ is minimal, and hence the exponent $\alpha$ takes a minimum value 
\begin{equation}
\alpha_\text{min}=\frac{2N}{N+1}
\end{equation}
\item At $g=2/N$ (which recalling that $g \in [-1,1]$ is only attainable for $N>1$), the exponent $\alpha=2$, dual to the non-interacting case.  We will look at this for the case $N=2$ in the next section.
\item In fact, $g=1/N$ is a self-dual point and there is a more general duality  (which as in the previous point is only relevant for $N>1$) 
\begin{equation}
g \rightarrow \frac{2}{N} -g
\end{equation}
under which the exponent $\alpha$ is invariant.
\item There are quantum phase transitions (QPT) where $d=1$ (corresponding to $\alpha \rightarrow \infty$) at 
\begin{equation}\label{g_qpt}
g_{c}=\frac{1\pm\sqrt{1+N}}{N}
\end{equation}
(see also Ref.~\onlinecite{Kiss2}).  For $g$ outside the range $(g_c^-,g_c^+)$, the hybridization is an irrelevant operator under RG, and hence the resonance width is zero, with a non-analytic jump in the occupation $n_d$ as $\varepsilon_0$ crosses $0$.  Considering once more that $g \in [-1,1]$, we see that there is a QPT for attractive interaction $U<0$ for any number of leads $N$; however there is only a QPT for repulsive interactions $U>0$ for $N \ge 3$.
\end{enumerate}

\begin{figure}\label{prefactor_ratios}
\begin{center}
\includegraphics[width=3in,clip=true]{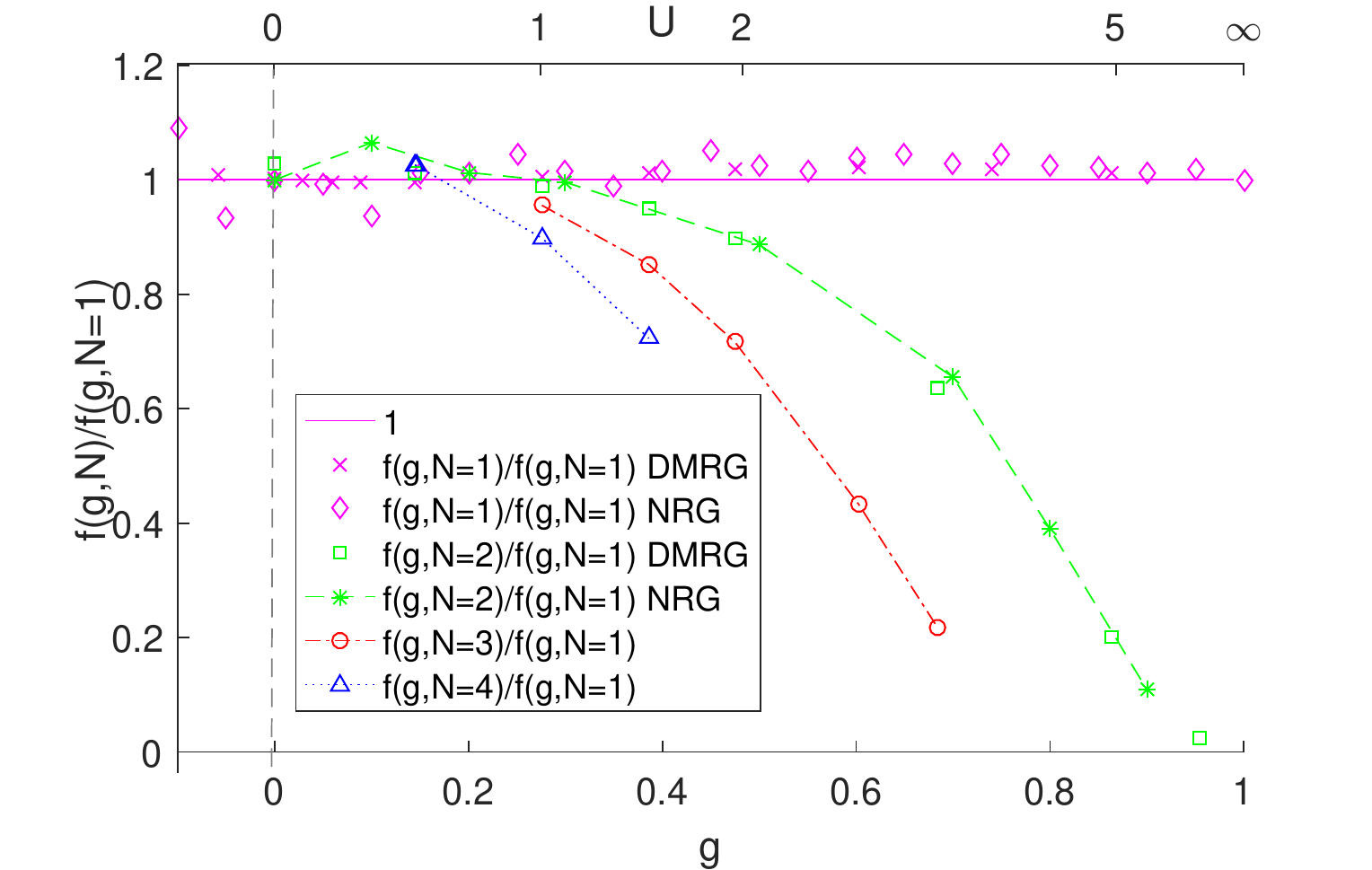}
\end{center}
\caption{A plot of $f_N(g)/f_1(\alpha)$, i.e. the ratio of the multichannel prefactors $f_N$ to the single channel case for the appropriate exponent $f_1(\alpha_N(g))$
obtained by DMRG. For $N>1$ lines have been included as a guide to the eye, showing the decay of the points as interaction is increased.}\label{fig:fNr}
\end{figure}

The third point is very curious.  It can be shown that in the bosonized description, one can make a linear combination of the $\phi$ fields so that the duality is exact -- i.e. the Hamiltonian maps onto itself.\cite{SchillerAndrei}  However, looking at the plot of the pre-factor, we see that this duality doesn't appear to be exact for any value of $N$.  This can be seen analytically for $N=2$ -- in this case going back to the definition of $g$ in Eq.~\eqref{defg}, we see that the duality corresponds to $\nu U \rightarrow 1/\nu U$.   Looking at the strong coupling limit $U\rightarrow \infty$ (which will be discussed in more detail), one indeed finds a mapping back onto the non-interacting model,\cite{SchillerAndrei}  but with a hybridization $t' \rightarrow (t/U)t'$ -- indeed this behavior of the pre-factor $f_{N}(g)\rightarrow 0$ as $g \rightarrow 1$ for $N=2$ can be seen in Fig.~\ref{fig:fN}.

For $N>2$, it is not so easy to do any analytic calculations as there is a phase transition (see the next section) before $g\rightarrow\infty$ and the interaction value dual to the non-interacting case is large but finite, outside the realm of either perturbation or strong coupling theory.  However, the numerical data in Fig.~\ref{fig:fN} clearly shows that this duality is not exact even for $N>2$, which has previously been questioned in Ref.~\onlinecite{Borda2}.

In the bosonized Hamiltonian, we can form an appropriate linear combination of fields $\phi_A=[(1-g)\phi_0-g\sum_{j=2}^N \phi_j]/\cal{N}$ where $\cal{N}$ is an appropriate normalization factor to retain the standard form of $H_0$ and the remaining fields are constructed to be orthonormal to each other and $\phi_A$.  This is analogous to the combinations of bosonic fields used in the conventional spin-charge separation.\cite{Gogolin,GiamarchiBook}  Under this transformation, the hybridisation term in the Hamiltonian Eq.~\eqref{eq:hybNtrans} takes on the same form as in the $N=1$ case, Eq.~\eqref{eq:bsG}. with $\beta^2=16\pi d$ with the scaling dimension $d$ given by \eqref{dual_eq}.  It would then appear that the pre-factor $f_N$ is given by the same expression as for the $N=1$ case, Eq.~\eqref{Gammaexpression}, with the appropriate exponent for the given value of $g$.  

While we have already seen that the breakdown in duality of the pre-factor means that this can't be correct, it is instructive to look at the ratio $f_N(g)/f_1(\alpha_N(g))$, which is plotted in Fig.~\ref{fig:fNr}.  The simplicity of this plot indicates that indeed the pre-factor calculated from the boundary sine-Gordon model, $f_1(\alpha)$ plays an important role in the overall pre-factor, however this is multiplied by something else that does not obey the duality and decreases monotonically with increasing interaction.  The theoretical origin of this extra factor is an open question, which we will discuss further in Section  \ref{sec:BAcomment2}.

\subsection{The line-shape $n_d(\varepsilon_0)$ for $N=2$}
\label{sec:ndN2}

\begin{figure}
\begin{center}
\includegraphics[width=3in,clip=true]{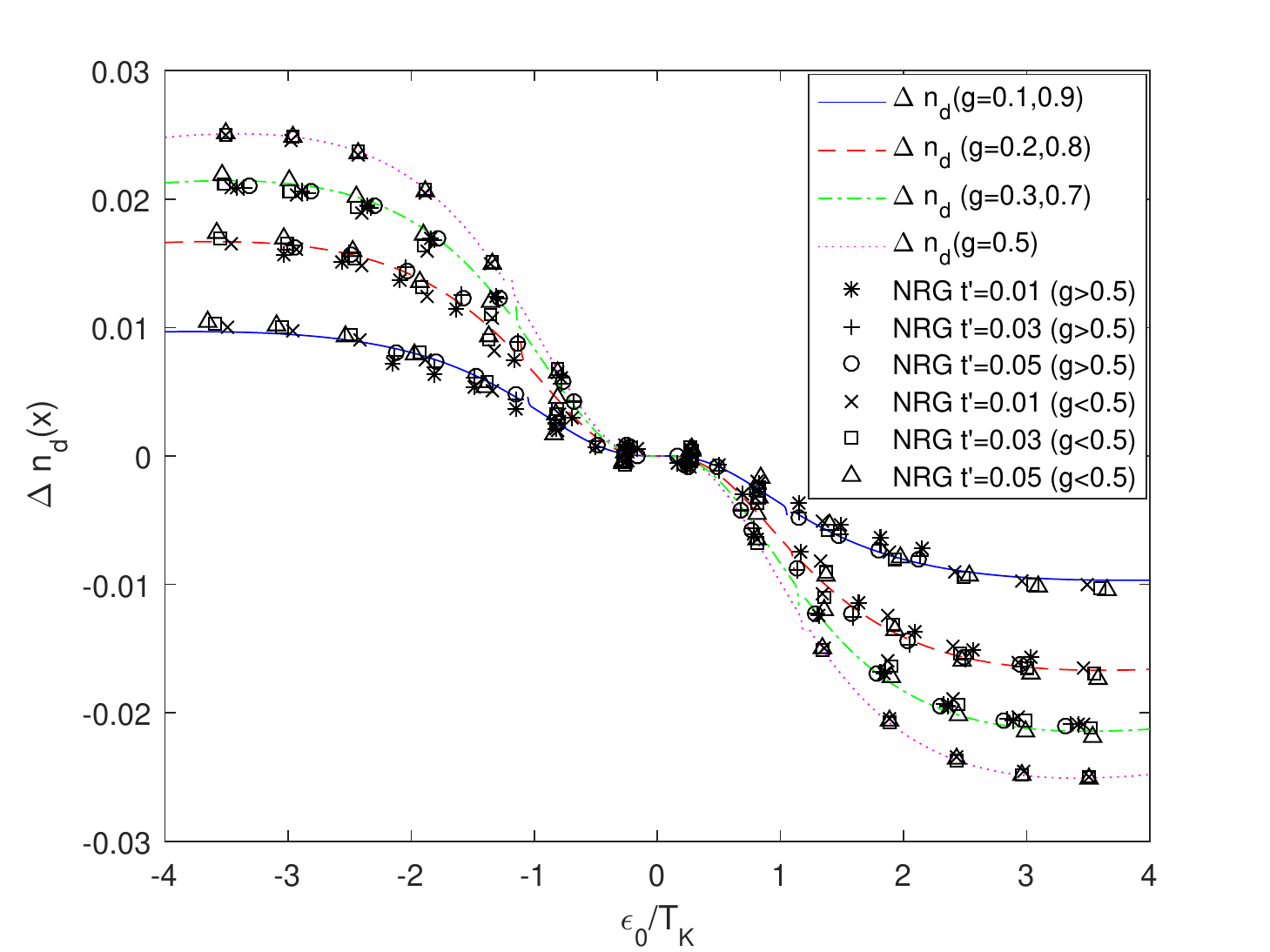}
\end{center}
\caption{The occupancy of the dot $n_d$ as a function of $x=\varepsilon_0/T_K$ for the two-channel case.  Like in Fig.~\ref{fig:nd}, the deviation from an arctan seen in the non-interacting case is rather small, so we have plotted the difference $\Delta n_d = n_d(g,x)-n_d(g=0,x)$.  The analytical line comes from Eq.~\eqref{nd_BA_equation} with the value of the exponent $\alpha$ given by Eq.~\eqref{alphaN}.}
\label{fig:ndN2}
\end{figure}

Let us now briefly go beyond the resonance width $T_K$ and look at the entire line-shape $n_d(\varepsilon_0)$ for the multichannel case.  Like in the single channel case discussed in Sec.~\ref{sec:BA} this is a function of the scaling variable $x=\varepsilon_0/T_K$, and is plotted for two leads $N=2$ for various values of $g$ in Fig.~\ref{fig:ndN2}.  There are two non-trivial observations one can make about this graph:
\begin{enumerate}
\item Unlike the resonance width $T_K$, the line-shape $n_d(\varepsilon_0/T_K)$ when expressed in terms of the scaling variable exhibits the duality $g \rightarrow 2/N \,-g$ discussed above, i.e. it is a unique function of the exponent $\alpha$.
\item When the interaction dependence is expressed in terms of the exponent $\alpha$, the occupancy $n_d(\varepsilon_0/T_K)$  is in fact given by exactly the same expression Eq.~\eqref{nd_BA_equation} as in the $N=1$ case.
\end{enumerate}
In other words, once one knows the exponent $\alpha$ and the emergent energy scale $T_K$, the number of leads $N$ \textit{does not} appear to play any further role in the thermodynamic properties of the model. Finally, let us point out the differences encountered when computing the value of $T_{K}$ for two different values of $g$ where the exponent $\alpha$ is the same. For instance, we take $g=0.2,0.8$, which are known to give a value $\alpha\sim 1.515$. In that case we find $T_{K}(g=0.2,t'=0.01)\approx 1.66\times 10^{-3}$ and $T_{K}(g=0.8,t'=0.01)\approx 4.89\times 10^{-4}$, showing that the prefactor $f_{N=2}(g)$ breaks the duality. At very strong interactions, and due to the presence of the prefactor $f_{N=2}(g)$ (see figure \ref{fig:fN}), the relevant energy scale of the problem becomes very small and its numerical extraction becomes more challenging, which might in turn induce more error when computing the line shape shown in Fig.~\ref{fig:ndN2}.

\subsection{Strong coupling for $N>1$}
\label{sec:QPT}

\begin{figure}
\begin{center}
\includegraphics[width=3in,clip=true]{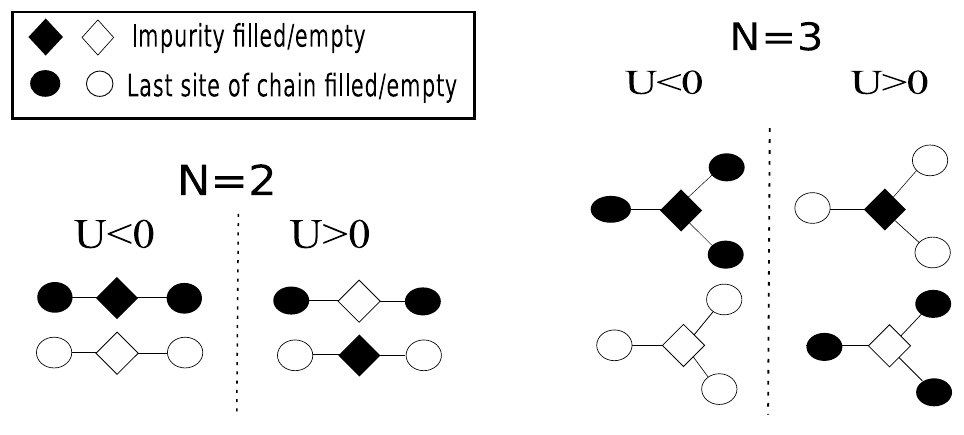}
\end{center}
\caption{Representation of the low energy subspace states in the strong coupling regime for $N=2$ and $N=3$ in the repulsive and attractive cases. The impurity site is represented by a square, whereas the site closest to the impurity on the tight-binding chain is represented by a circle. The legend shows whether these sites are filled (black) or empty (white).}
\label{fig:strong_N}
\end{figure}

We now discuss the strong coupling limit $U\rightarrow \pm \infty$ for the case $N>1$.  As in the $N=1$ case described in Sec.~\ref{sec:strong}, one can think of an enlarged impurity consisting of the impurity and the final lattice site of each of the chains (making a total of $N+1$ sites).  For each value of $N$ and sign of $U$, there are then exactly two low-energy states of this enlarged impurity, with the remaining states separated by an energy of $|U|$.  This is represented pictorially for $N=2$ and $N=3$ in Fig.~\ref{fig:strong_N}.  The effective (enlarged) impurity is then hybridised with the $N$ leads (each one missing their last site), so like the original weak coupling problem, the strong coupling limit is a two-level system coupled to $N$ leads.  The question is about what the effective couplings to the leads are when everything is projected into the low-energy subspace.


Consider the attractive regime $U<0$ first. 
As seen in Fig.~\ref{fig:strong_N}, the two low-energy states of the effective impurity are either all empty or all filled -- hence the two states differ by $N+1$ particles.  In order to get $N+1$ fermions onto (or off) the impurity from $N$ leads, at least two of the fermions need to jump from (to) the same channel. The scaling dimension of this effective hybridisation  operator is therefore:\cite{Gogolin}
\begin{eqnarray}
d=2 + \frac{N-1}{2}>1
\end{eqnarray}
As a boundary operator, this should be compared to $1$, where it is clear that $d>1$ always, for any number of leads (including $N=1$).  Hence we conclude that in the $U<0$ case, the effective hybridisation between the effective enlarged impurity and the leads is an irrelevant operator.  Hence the level width remains zero, with a discontinuous jump in occupation at $\varepsilon_0=0$.  Contrasting this to the small (negative) $U$ case where the hybridisation is relevant, the resonance has a finite width, and there are no discontinuities in the occupation as a function of $\varepsilon_0$, we see that at some finite $U<0$, there must be a quantum phase transition between the two states of the system.\cite{Kiss2,Borda1}  This is in complete agreement with our earlier bosonization analysis, Eq.~\eqref{g_qpt}.



The repulsive case, $U>0$ is a bit different.  In this case, the difference in number of particles between the two different low-energy states is $N-1$, as seen in Fig.~\ref{fig:strong_N}.  We have already analysed the interesting case $N=1$ in Sec.~\ref{sec:strong} where no hybridisation to the leads to make a transition between the two states is required to leading order; we have also already discussed the case $N=2$ where the effective hybridisation to the leads is a single particle hopping, thus mapping the IRLM back onto itself, albeit with a hybridisation suppressed by a factor $t/U$ from second-order perturbation theory (see also Ref.~\onlinecite{SchillerAndrei}.  For general $N$, we can again write an effective hybridisation operator between the leads that hops $N-1$ particles on to (off) the impurity; with these distributed between $N$ leads, no two have to go onto the same lead.
Hence the scaling dimension of an operator of this type is:
\begin{eqnarray}
d=\frac{N-1}{2}
\end{eqnarray}
Again, as a boundary operator, this should be compared with $d=1$.  This means that the case $N=3$ represents the marginal case $d=1$, with the phase transition taking place exactly at $U=+\infty$ (again consistent with the bosonization description,  Eq.~\eqref{g_qpt}), whereas for $N>3$ the operators describing the virtual process are irrelevant meaning there is a phase transition at finite $U>0$.  This is also the reason why in the cases $N=1$ and $N=2$ no phase transition is encountered in the repulsive region ($U>0$), since the low energy effective hamiltonian is always described by relevant operators, whereas for $N\geq 3$, there is a specific value of $g$ from where low energy physics are described by irrelevant operators.\\

\subsection{The Bethe Ansatz solution revisited}
\label{sec:BAcomment2}

Let us now return to our discussion of the Bethe Ansatz solution of the IRLM that we begin in Sec. \ref{sec:BAcomment}.  From our exact expression for the exponent, \eqref{alphaN}, we see that the dependence on the number of leads $N$ comes in the quadratic term $Ng^2$ only.  However, results found from the Bethe Ansatz (summarised in Table~\ref{tab:exponents}) do not find a quadratic term.  Indeed, in Ref.~\onlinecite{Ponomarenko}, the model was analysed for general $N$, with no dependence on $N$ found in the results.  It is known that in integrable field theories, the regularisation procedure can lead to different results, particularly with regard to extracting non-universal exponents from bare parameters (see e.g. Ref.~\onlinecite{Saleur1999}).  Indeed, we have three different regularisation procedures in this work: that of the fermionic field theory, that of the bosonized field theory, and that of the lattice used in the numerics.  However, we feel that in this particular case, this is not the relevant issue.  

Indeed, there is an issue of short-distance regularisation which we have shown can be exactly treated in this model by using the phase-shift as the parameter rather than the bare interaction -- and the peculiar feature of this model having the interaction only acting over the weak link means that the interacting model can be solved exactly for $t'=0$ -- giving an analytic link between bare parameters in any regularization and the phase shift that enters the many-body calculation.  Furthermore, we have shown that in the $N=1$ case, both field theory and the lattice give the same answers, so long as this short-distance behavior has been taken into account in the correct way.\cite{gonzalo_phd}   In addition, the agreement to numerics for the dot occupancy from the Bethe ansatz solution is too good to be a coincidence.  It is therefore far from clear why the Bethe ansatz solution fails to get the exponent correct, while it does agree with numerics if a renormalization of the exponent is forced by hand.  It is even more telling that the direct Bethe ansatz gets an exponent that is independent of the number of leads, as particularly in thinking about the strong coupling limits, it is very difficult to imagine a regularisation where this should be the case.

Taking this point further, if we look in detail at the multi-channel Bethe Ansatz solution,\cite{Ponomarenko} we see that very much like Eq.~\eqref{HlatticeN}, only one channel plays an important role in the two-body wavefunction (and hence the $S$ matrix) of the model, with the remaining $N-1$ channels (which do not have a hybridization with the dot) playing a passive role.  It is clear however from our solution that these channels play an important role in determining the scaling dimension of the hybridisation operator, which in turn must come from fluctuations in local density in these channels.  This shows that the filling of the states, and the consequent dressing of the Bethe equations are likely to play an important role in getting the correct solution.  Indeed, the $S$ matrix for the IRLM is equivalent to that of the massless-limit of the massive Thirring model,\cite{TsvelickWiegmann} which is known to be equivalent to the sine-Gordon model\cite{SColeman} which appears at the level of the Bethe equations through dressing the $S$ matrix due to the filled Fermi sea,\cite{Korepin,Korepin2} although it is worth pointing out that much less has been done in this direction on the boundary sine-Gordon model used in this work.  While such an equivalence may explain the success of our boundary-sine-Gordon model analysis in Sec.~\ref{sec:prefactor}, a direct calculation of this using the Bethe ansatz is an open question.

It is also worth commenting that once this renormalization of the parameters of the $S$-matrix takes place, the calculations from the thermodynamic Bethe ansatz appear to be completely consistent with numerical results, even for the multi-lead case, as demonstrated in Sec.~\ref{sec:ndN2}.  It would be an interesting exercise to see if such a renormalization would also give agreement with numerical results for non-equilibrium properties.\cite{Andrei}

\section{Summary}

A complete thermodynamic theory on the IRLM Eq.~(\ref{Hlattice}) has been presented in this work, showing very good agreement between exact analytical expressions and numerics. We have proved very constructive to analyze the single channel version of the model first in order to extrapolate results to the (general) multichannel version. 

The problem of matching numerical and analytical results in the model\cite{Borda1} for $U\nu\gg 1$ has been identified to be twofold: First, that is the \emph{phase shift} $\delta$ Eq.~(\ref{phase}) and not $U$ the relevant parameter of the theory to be considered. This scattering phase shift, which appears in the single particle problem of $t'=0$, allows to study the model for \emph{any} desired value of interaction $U$, even if these values are way above the bandwidth of the bulk. The substitution of the bare parameter $U$ by $\delta$, the \emph{phase shift} substitution, makes a clear connection between the original microscopic model in the theory at $t'=0$ and the equivalent (weakly-coupled $t'/t<<1$) treatment when fluctuations on the dot are allowed. In turn, the phase shift substitution introduces the interaction parameter $g$ given by (\ref{defg}) into the problem, which allows tto study the model in the whole range of interactions $U$. Secondly; the perturbative RG treatment gives the correct scaling of $t'$ with the interaction parameter $g$, thus providing the correct asymptotic form of the relevant energy scale $T_{K}$ eq.(\ref{exp_asymptotic}). This scale is dominated by a single exponent $\alpha$ that depends on the interaction coupling $g$ as given by eq.~(\ref{alpha}), which has been confirmed by NRG and DMRG in section III.C. In addition, we have shown the exact form of $T_{K}$ to include an interaction dependent prefactor that must be taken into account when comparing with numerics.  The prefactor is identified from a well known integrable theory with a boundary term,\cite{FLSaleur} the Boundary sine-Gordon model (BsG). This correspondence between the lattice model with microscopic parameters (the IRLM) and the BsG has been checked numerically by both NRG and DMRG techniques.

In order to confirm recent integrability results in the model, the exact solution of the IRLM via Bethe-ansatz as given in Ref.~\onlinecite{Rylands} has been investigated in section IV. We have shown expressions to be in very good agreement with NRG numerics if the thermodynamic exponent $\alpha$ is used in the form of eq.~(\ref{alpha}), which is obtained directly from Bosonization. We conclude the exact calculation of the line shape $n_{d}$ to be in excellent agreement with performed NRG numerics, confirming universality of $n_{d}(x)$, where $x=\varepsilon_{0}/T_{K}$ is the scaled variable. We show that in order to see clear separation between different line shapes at different interactions $g$, the non-interacting form of $n_{d}$ must be subtracted. Finally, we emphasize that the Bethe-ansatz method does not reproduce \cite{Rylands,FilyovW,Ponomarenko,TsvelickWiegmann} 
this form of the exponent $\alpha$ as it is obtained by Bosonization. In this sense, a careful check of the calculation for $\alpha$ by using the Bethe-ansatz is strongly desirable.

Once all details of the single channel IRLM are known, its extension to the $N$ channel version has proved to be quite straightforward, although we want to make some observations here. In particular, we have shown the exact results computed in Ref.~\onlinecite{Rylands} for the occupation of the dot to hold fairly well for the $N=2$ case, by just using the appropiate form for $\alpha$ (Eq.~\eqref{alphaN}) and $T_{K}$ (Eq.~\eqref{TKexactN}). Supported by the NRG simulations in the $N=2$ case, we believe relation (\ref{nd_BA_equation}) to hold for any number of leads $N$ providing $\alpha$ and $T_{K}$ are given by Eqs.~\eqref{alphaN} and \eqref{TKexactN} respectively.

In order to verify the expression for the exponent, Eq.(\ref{alphaN}) as obtained by Bosonization, the DMRG technique was extended to up to four leads, resulting in very good agreement between the numerically obtained exponents and the analytic expression. It is important to note that the exponent $\alpha$ for $N$ leads differs from the $N=1$ case only in the $g^{2}$ term, and that for $N>1$, such exponent always presents a \emph{duality} between different $g$ regions. For $N=2$ channels, which is the relevant model for transport,\cite{Andrei,Vinkler,Borda1,SchillerAndrei,Boulat1,Boulat2,BBSS,Sampaper,Sampaper2,Sampaper3,vonDelftIRLM2018}
this duality relates the $U\leftrightarrow 1/U$ sectors, and it would appear in principle that a weak-to-strong coupling correspondence is always present in the $N=2$ case. We check this dual relation for the $N$ channel IRLM in section IV.A, showing that the prefactor of $T_{K}$ breaks this duality in the energy scale, in accordance with what is obtained from a strong coupling expansion in the lattice.\cite{SchillerAndrei} We hint a possible relation of the $N\neq 1$ prefactor with the $N=1$ case (Fig.~\ref{fig:fN}) guided by DMRG ($N=1,2,3,4$) and NRG ($N=1,2$) numerics, although the exact calculation of such prefactor from bosonization is at the moment unknown to us. The only reliable result we are giving here is that the prefactor of the $N>1$ IRLM does not solely depend on $\alpha$, this being the cause of the duality breaking in $T_{K}$. We stress that this result does not appear to affect universal properties of the model like the curve $n_{d}$, which includes $\alpha$ as the only parameter, therefore conserving the duality $U\leftrightarrow1/U$ in the $N=2$ case, which we have checked by NRG in section VI.B.

\section{ACKNOWLEDGEMENTS}

The authors want to thank H.~Saleur and J.~Quintanilla for useful comments on the work presented here. G.~Camacho acknowledges a 50th anniversary scholarship from the University of Kent. P.~Schmitteckert was supported by ERC-StG-Thomale-TOPOLECTRICS-336012.  We are grateful to the Rechenzentrum WŸrzburg for providing computational resources through the DFG funded compute server Julia, INST 93/878-1, of the University of W\"urzburg.


\appendix
\section{Some numerical details}
\label{sec:numapp}

\subsection{NRG}
\label{sec:NRG}
The Numerical Renormalization Group (NRG) \cite{Wilson:RMP75,Krishnamurthy_Wilkins_Wilson:PRB1980} allows to compute low-energy properties of a one dimensional system by construction of an effective tight-binding hamiltonian. The method relies on a logarithmic discretization of the band (the bulk's density of states), where such a discretization is controlled by the parameter $\Lambda$. The chosen value of $\Lambda$ depends on the system under consideration, but typically $\Lambda=1.5 - 2.5$. The bigger the value of $\Lambda$, the smaller the size of the chain we need to use to capture low-energy features. On the other hand, the continuum limit (or infinite chain) of the model is recovered when $\Lambda\to 1$. This logarithmic discretization of the band makes possible to map the original hamiltonian into a lattice description, where the hopping amplitudes between neighbouring sites acquire a dependence on $\Lambda$. Concretely, for the IRLM, the effective hamiltonian after logarithmic discretization is\cite{Costi}
\begin{eqnarray}
H=\sum_{n=1}^{N}t_{n}c_{n+1}^{\dagger}c_{n}+\text{h.c} + \varepsilon_{n}c_{n}^{\dagger}c_{n}+H_{\text{imp}}\nonumber\\
H_{\text{imp}}=\varepsilon_{0}d^{\dagger}d + t'(d^{\dagger}c_{0}+\text{h.c})+U:d^{\dagger}d::c_{0}^{\dagger}c_{0}:
\end{eqnarray}
Here the notation $:A:=A-1/2$. The dependence of $t_{n}$ with the discretization parameter has, for a constant denssity of states in the bulk, the following form\cite{Wilson:RMP75,Krishnamurthy_Wilkins_Wilson:PRB1980,Costi}:
\begin{eqnarray}
t_{n}=\frac{1}{2}(1+\Lambda^{-1})\Lambda^{n/2}
\end{eqnarray}
The on-site energies $\varepsilon_{n}=0$ within a good approximation\cite{Costi}. The part $H_{\text{imp}}$ refers to the system composed by the impurity site and the neighbouring site of the chain. One starts by diagonalizing this part, computes the eigenvectors, and then adds an extra site to the chain, repeting the diagonalization procedure. The general iterative algorithm for the $N+1$ step is given by\cite{Costi}:
\begin{eqnarray}
H_{N+1}=\sqrt{\Lambda}H_{N} + \Lambda^{N/2}t_{N}c_{N+1}^{\dagger}c_{N}+\text{h.c}
\end{eqnarray}
where we have taken $\varepsilon_{n}=0$ for all sites.\\

The dot occupation is given by averaging the number operator at the impurity over the ground state of the system:
\begin{eqnarray}
n_{d}=\langle GS|d^{\dagger}d |GS\rangle
\end{eqnarray} 
Thus, the operator needs to be computed at every iteration step. We are interested in computing values of $n_{d}(\varepsilon_{0})$, that is, as a function of the resonant level energy. The reason to do this is to compute the value of the width at any interaction $U$:
\begin{eqnarray}\label{Tk_appendix}
T_{K}^{-1}=-\pi\bigg(\frac{\partial n_{d}}{\partial\varepsilon_{0}}\bigg)_{\varepsilon_{0}=0}
\end{eqnarray}

 In doing so, we must take care to stay within a range of $\varepsilon_{0}$ sufficiently small, so that the slope is taken by linear fitting approximation. By computing the value of $T_{0}$ (that is $T_{K}=0$), one can calculate the effective bulk density of states parameter $\nu$ of the theory:
\begin{eqnarray}\label{bulk_rho}
 T_{0}=\pi\nu(t')^{2}
\end{eqnarray}
The value of this parameter $\nu$, which is calculated numerically, is then used in the analytical formulas (eq.(\ref{TKexact})). The use of the numerically obtained $\nu$ as opposed to the theoretical value $\nu=1/\pi t$ is a key step in order to see full match between numerics and analytical formulas.\\

Once the energy scale (\ref{Tk_appendix}) is known for different values of $t'$ and a fixed value of $g$, the prefactor $f(g)$ eq.(\ref{Gammaexpression}) and the exponent $\alpha$ eq.(\ref{alpha}) can be calculated by linear fitting; that is, representing $\log(T_{K})$ vs $\log(t')$. For all simulations developed in section III.C, different values of $\Lambda$ and the total number of sites were used. In particular, for $\Lambda=1.5$ a total of $S_{k}=500$ states were kept under truncation of the hamiltonian, for a total size of $N=82$ sites on Wilson's chain. For $\Lambda=1.3$, more states are needed, therefore we chose $S_{k}=1000$ states under truncation, for a total size of the chain of $N=122$ sites. Each simulation is then repeated for several values of $\varepsilon_{0}$ in order to get the line shape for a fixed value of $U$, therefore allowing to calculate $T_{K}$ in (\ref{Tk_appendix}).

\subsection{DMRG}
\label{sec:DMRG}

While the NRG method is the standard one for interacting impurity problems, it relies on
the hypothesis of a separation of energy scales.\cite{Wilson:RMP75,Krishnamurthy_Wilkins_Wilson:PRB1980}
In order to verify the NRG simulations
we applied the DMRG, which has the advantage of including a back feed from low to high energy scales,
which is missing in the NRG. For this reason the DMRG does not rely on a separation of energy scales.
The disadvantage of the DMRG compared to NRG is that it is 
significantly more expensive computationally.

\begin{figure}
\begin{center}
\includegraphics[width=\columnwidth,clip=true]{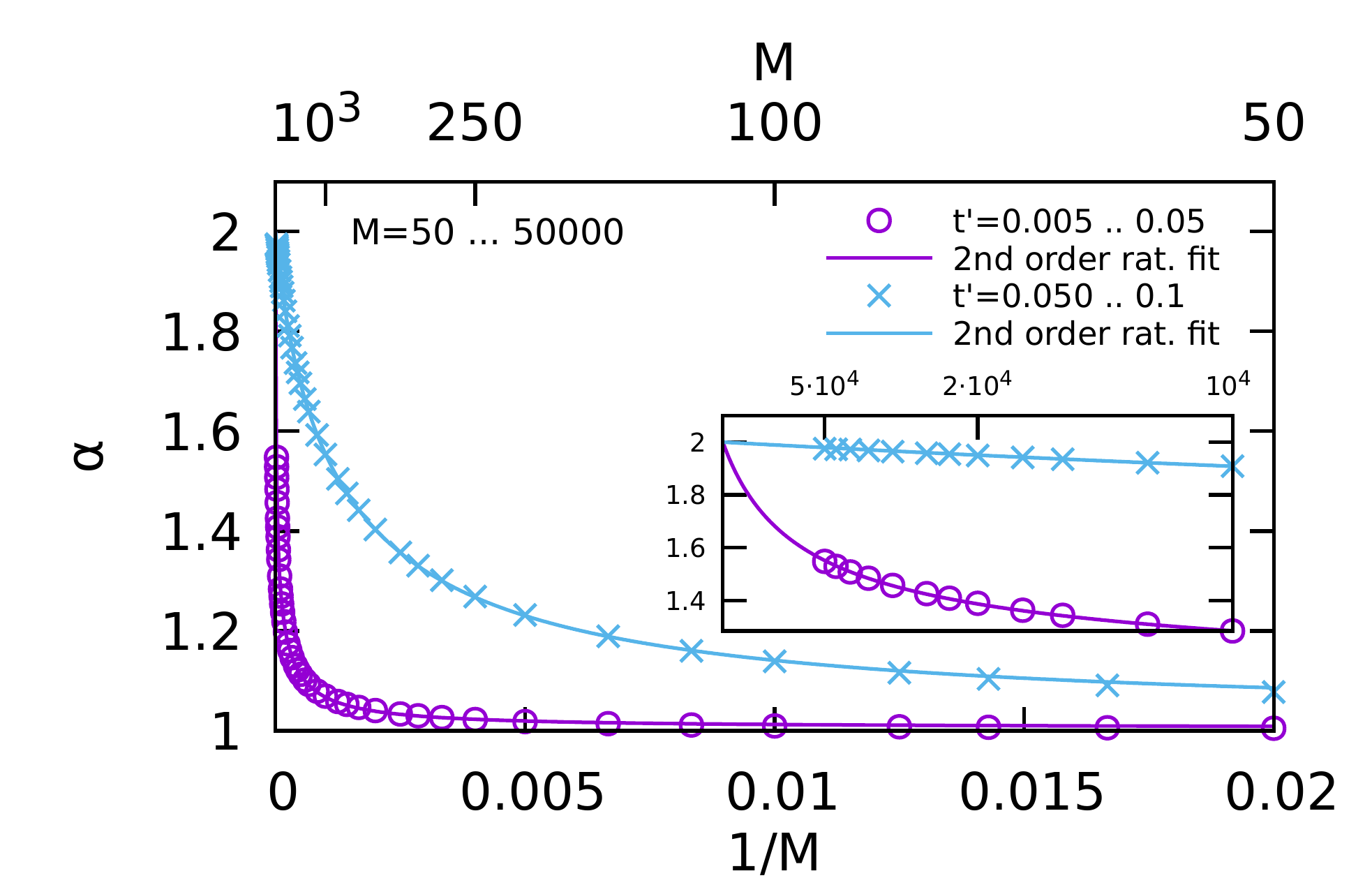}
\end{center}
\caption{Scaling exponent $\alpha$, Eq.~\eqref{exp_asymptotic} for the non-interacting case of model \eqref{Hlattice}
as obtained from diagonalizing the quadratic form for $M$ sites, one for the impurity and $M-1$ for the lead.
The scaling exponent $\alpha$ is obtained by fitting a power law to the susceptibility in the range
$t' \in [ 0.005, 0.05]$ (circles) $t' \in [ 0.05, 0.1]$ (crosses) for system sizes $M$ ranging 
from 50 to $50\,000$.
The lines represent a fit to a second order rational function.  The inset zooms in on the larger system sizes showing more clearly the extrapolation to an infinite system size with $\alpha=2$.}
\label{fig:fs1}
\end{figure}

\begin{figure}
\begin{center}
\includegraphics[width=\columnwidth,clip=true]{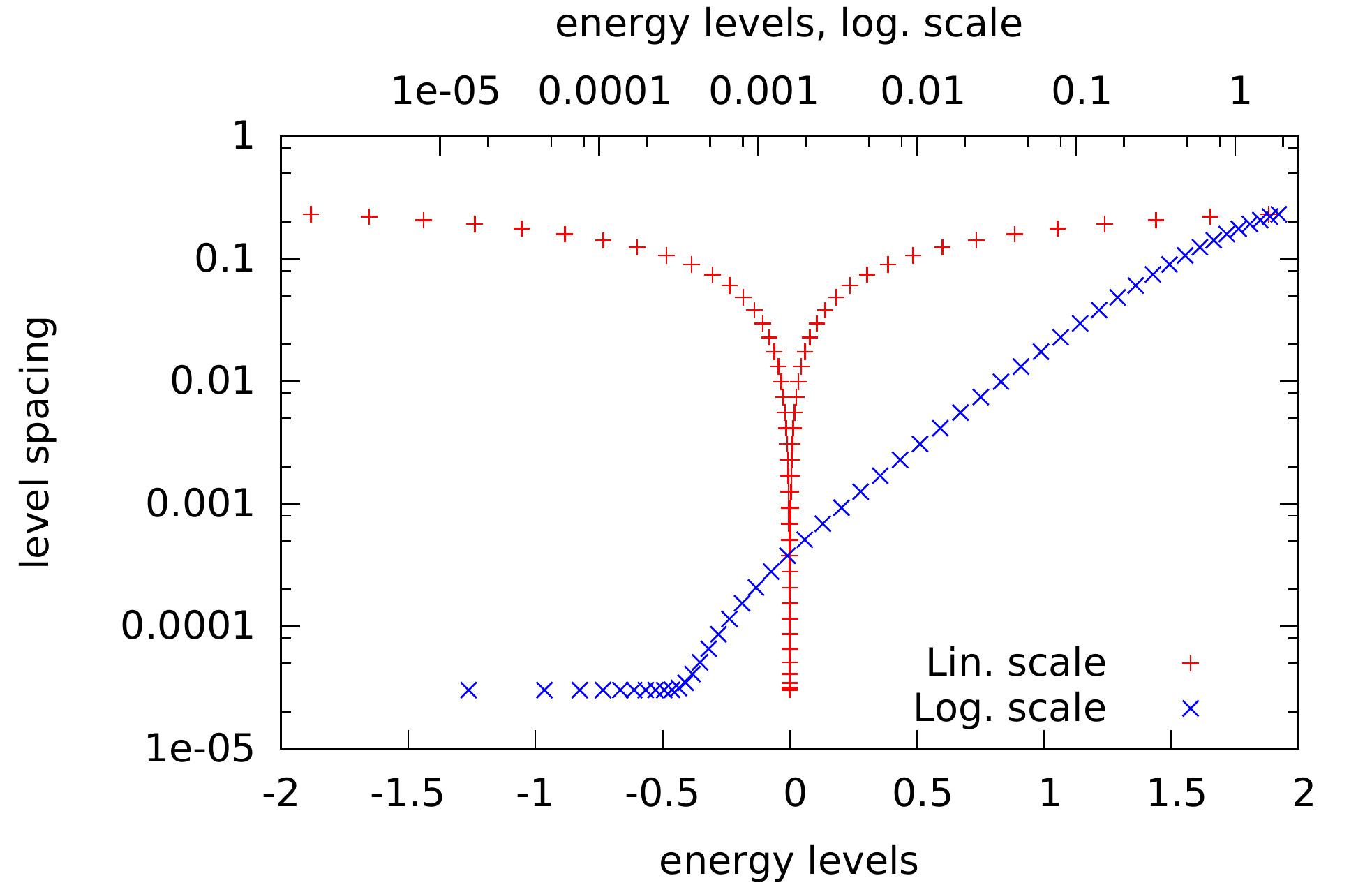}
\end{center}
\caption{Level spacing of the distribution of the energy levels of the $M_\mathrm{L}=100$ site leads.
The  pluses show the position of the energy levels on a linear scale, lower axis,
while the crosses  correspond to the positive levels on logarithmic scale, top axis.
The distribution of the levels is particle hole symmetric.  }
\label{fig:fs2}
\end{figure}

Before giving some details of the DMRG \cite{White:PRL92,White_Noack:PRL92,White:PRB93} procedure used in this work, 
let us mention briefly a slightly surprising feature of the resonant level model (see also Ref.~\onlinecite{Saleur_PS_Vasseur:PRB2013}).  
It turns out that if one naively simulates the model \eqref{Hlattice} on a lattice, even for the non-interacting case, 
one finds very strong finite-size effects.  
For example, in Fig.~\ref{fig:fs1} the exponent $\alpha$ which should be $2$ is numerically extracted for the non-interacting case with a tight binding lead of size $M-1$.  It is seen that even for $M=20\,000$, one finds an exponent of $\alpha \approx 1.95$ or $\alpha \approx 1.39$ 
depending on the values of $t'$ used for fitting the power law, far below the true value.  
By performing finite size scaling, one can indeed extrapolate to the infinite system size limit and reproduce $\alpha=2$, 
however this is very expensive numerically as one still has to go to large system sizes.

To get around this problem, one may apply Wilson chain like leads in NRG, which are called damped \cite{BohrSchmitteckertWoelfle:EPL2006,Schmitteckert:JPCS2010}
or smoothed \cite{Vekic_White:PRB1993} boundary conditions in the context of DMRG.
In this work we represented the leads in energy space as described in Refs.~\onlinecite{BohrSchmitteckert:PRB2007,Schmitteckert:JPCS2010}
and displayed in Fig.~\ref{fig:fs2}. To this end we coupled the impurity to $n$ sites representing the first
site of each of the $n$ leads. We then added a few additional sites in real space in order to keep the total number
of real space sites even, specifically we used 2 (n=1), 8 ($n=2,3$), and 10 ($n=4$) real space sites in total
and 100 ($n=1,2$), 50 ($n=3,4$) sites in energy space for each lead with a linear dispersion relation and a band cut-off of 2.
In addition we applied other discretization schemes to ensure that our results are not spoiled by discretization effects.

Following Refs.~\onlinecite{BohrSchmitteckert:PRB2007,Schmitteckert:JPCS2010} we distributed the levels on logarithmic scale, with the exception of a few level closest
to the Fermi points, where we applied a linear discretization and coupled the last site of
each lead in real space to a lead in energy space. For a detailed discussion concerning the representation
in energy space we refer to Ref.~\onlinecite{Schmitteckert:JPCS2010}.

Although the leads in energy space allow for a high energy resolution of the leads,
and therefore of the physics, they are at risk 
of getting trapped to an excited state within DMRG.\cite{PS:Proceedings98}
In order to avoid this problem we applied the sliding block $B$ approach of Ref.~\onlinecite{Schmitteckert:PRB2018}.
We kept enough states to ensure a discarded entropy, the entropy of the information
thrown away in a DMRG step, is below $10^{-9}$ and for most of the steps significantly smaller.
In addition we performed scaling sweeps in the spirit of \cite{BohrSchmitteckert:PRB2007} by first solving
the problem for a rather large coupling of $t'=0.4$ and $\epsilon_0=0$.
We then performed DMRG runs by restarting and successively lowering  $\epsilon_0$.
In addition we kept the restart files for $\epsilon_0=0$ and restarted it for a smaller $t'$.
We kept on performing $\epsilon_0$ sweeps while restarting the $\epsilon_0=0$ runs in order to lower $t'$.
Following this procedure we ensure that we do not converge to an excited state.

\section{Some technical notes on the Bethe-Ansatz expression for $n_d(\varepsilon)$}
\label{sec:appBA}

Let us compare the general expression for $n_d(\varepsilon)$, Eq.~\eqref{nd_BA_equation}, to the known limits in the non-interacting $\alpha=2$ and strongly interacting $\alpha=1$ cases.  We define the scaling variable $x=\varepsilon/T_B$, and for technical reasons, it is slightly easier to consider the derivative, which is a generalised susceptibility.
\begin{multline}
\chi(x) = -\frac{d n_d}{dx} \\
= \frac{1}{\sqrt{\pi}}\sum_{n=0}^{\infty}\frac{(-1)^{n}}{n!}\frac{\Gamma\big(1+\frac{\alpha}{2}(2n+1)\big)}{\Gamma\big(1+\frac{\alpha-1}{2}(2n+1)\big)}
  x^{2n} 
  \label{eq:chix}
\end{multline}

\noindent \textbf{Non-interacting case}: substituting $\alpha=2$ into Eq.~\eqref{boundary_temp_relation} gives us the relationship $T_B=2 T_K$ so $x=\varepsilon_0/2T_K$, which can then be inserted into the non-interacting expression Eq.~\eqref{nd-U0} and differentiated to get the standard Lorenzian form of the susceptibility
\begin{equation}
\chi(x) = \frac{2}{\pi} \frac{1}{1+(2x)^2}.
\end{equation}
On the other hand, substituting $\alpha=2$ into the general expression Eq.~\eqref{eq:chix} gives
\begin{equation}
\chi(x) =  \frac{1}{\sqrt{\pi}}\sum_{n=0}^{\infty}\frac{(-1)^{n}}{n!}\frac{\Gamma\big(2+2n\big)}{\Gamma\big(3/2+n\big)}
  x^{2n} 
\end{equation}
Now using standard properties of Gamma functions, $\Gamma(2+2n)=(2n+1)!$, $\Gamma(3/2+n)=\sqrt{\pi}\, 1.3.5\ldots(2n+1)/2^{n+1}$, and $n! = 2.4.6\ldots2n/2^n$.  Hence
\begin{equation}
\chi(x) =  \frac{2}{\pi}\sum_{n=0}^{\infty} (-1)^{n} (2x)^{2n} = \frac{2}{\pi} \frac{1}{1+(2x)^2}\end{equation}
as required.

\noindent \textbf{Strong coupling case}: substituting $\alpha=1$ into Eq.~\eqref{boundary_temp_relation} gives us the relationship $T_B=\pi T_K/2$ so $x=2\varepsilon_0/\pi T_K$.  Combining this with Eq.~\eqref{eq:TKg1} that says $T_K=4t'/\pi$ in this case gives us $x=\varepsilon_0/2t'$.
Hence from Eq.~\eqref{occupation_x}, we have
\begin{eqnarray}
n_{d}(x) &=& 1-\frac{1}{1+(x-\sqrt{1+x^{2}})^{2}} \nonumber\\
&=& 1 - \frac{1}{2\left(1+x^2-x\sqrt{1+x^2}\right)}
\end{eqnarray}
Now, multiplying the top and bottom of the fraction by $1+x^2+x\sqrt{1+x^2}$ gives a much simpler expression
\begin{equation}
n_d(x) = \frac{1}{2} \left( 1 - \frac{x}{\sqrt{1+x^2}} \right)
\end{equation}
which can be differentiated to give
\begin{equation}
\chi(x)=\frac{1}{2}(1+x^2)^{-3/2}
\label{eq:chixa1}
\end{equation}
Now, we can substitute $\alpha=1$ into the general series Eq.~\eqref{eq:chix} and manipulate the Gamma functions to get
\begin{multline}
\chi(x) = \frac{1}{\sqrt{\pi}}\sum_{n=0}^{\infty}\frac{(-1)^{n}}{n!}\frac{\Gamma\big(3/2+n\big)}{\Gamma(1)} x^{2n}  \\
 =\frac{1}{2}+\frac{1}{2}\sum_{n=1}^{\infty} \frac{(-3/2).(-5/2)\ldots.(-2n-1)/2}{n!} x^{2n}
\end{multline}
which is the power series for the otherwise obtained expression Eq.~\eqref{eq:chixa1}.  Thus we have proved that the general expression in the main text Eq.~\eqref{nd_BA_equation} matches the known analytic results in both the non-interacting and the strongly-interacting limits.

\noindent \textbf{Large $\epsilon_0$:} For completeness, we also write the complementary series which is adapted from Refs.~\onlinecite{Ponomarenko,Rylands} and is needed to plot $n_d(\varepsilon)$ in Fig.~\ref{fig:nd}:
\begin{equation}
n_d(\varepsilon_0) = \frac{1}{2\sqrt{\pi}} \sum_{n=1}^\infty \frac{(-1)^{n+1}}{n!} \frac{\Gamma(1/2 + n/\alpha)}{\Gamma(1 - \frac{\alpha-1}{\alpha}n)} \left( \frac{\varepsilon_0}{T_B} \right)^{-2n/\alpha}.
\end{equation}
The crossover from one series to the other is at $\varepsilon/T_K \approx 1$.

\section{Coupling to a Luttinger Liquid}
\label{sec:appLL}

On coupling to a Luttinger liquid, the bosonized Hamiltonian \eqref{eq:H0bos} of the IRLM becomes
\begin{eqnarray}
H_{0}=\frac{K}{2}\int_{-\infty}^{+\infty} dx(\partial_{x}\phi(x))^{2}+\sqrt{\pi} g \; \partial_{x}\phi(0)S^{z}
\label{eq:H0bos}
\end{eqnarray}
where $K$ is the Luttinger liquid parameter, and $K=1$ corresponds to the non-interacting leads.

On making the scale change $\phi \rightarrow \sqrt{K} \phi$, we arrive at the full Hamiltonian (c.f. Eq.~\ref{hambos})
\begin{eqnarray}
H=H_{0}+\frac{t'}{\sqrt{2\pi}}\big(\eta_0 \eta e^{-i\sqrt{4\pi/K}\phi(0)}S^{-}+\text{h.c}\big)+\varepsilon_0 S^z
\label{hambosLL}
\end{eqnarray}
where 
\begin{eqnarray}
H_{0}=\frac{1}{2}\int_{-\infty}^{+\infty} dx(\partial_{x}\phi(x))^{2}+\sqrt{\pi/K} g \; \partial_{x}\phi(0)S^{z}
\label{eq:H0bosLL2}
\end{eqnarray}
Making the unitary transformation
\begin{eqnarray}
\bar{H}=\mathcal{U}^{\dagger}H\mathcal{U}
\end{eqnarray}
with
\begin{eqnarray}
\mathcal{U}=e^{i\sqrt{4\pi} g S^{z}\phi(0)}
\end{eqnarray}
eliminates the interaction term to give
\begin{multline}
\bar{H}=\frac{1}{2}\int_{-\infty}^{+\infty} dx(\partial_{x}\phi(x))^{2} \\
+\frac{t'}{\sqrt{2\pi}}\big(S^{-}e^{i\sqrt{4\pi/K}(1-g)\phi(0)}+\text{h.c}\big)
\label{eq:vertexLL}
\end{multline}
The scaling dimension of the vertex operator is then $d=(1-g)^2/2K$ as advertised in the main text.

\bibliographystyle{apsrev4-1}
\bibliography{PAPER_BIBLIOGRAPHY}

\end{document}